\documentclass[12pt]{article}

\usepackage{amsmath,psfig,epsf,epsfig}

\def\eqn#1{Eq.~#1}
\def\eqns#1{Eqs.~#1}
\def\sec#1{Sec.~#1}
\def\fig#1{Fig.~#1}

\begin{document}

\title{Self-gravitating Brownian systems and bacterial
populations with two or more types of particles.}

\author{Julien Sopik, Cl\'ement Sire and Pierre-Henri 
Chavanis}


\maketitle

\begin{center}
Laboratoire de Physique Th\'eorique (UMR 5152 du CNRS), 
Universit\'e Paul Sabatier,\\ 
118, route de Narbonne, 31062 Toulouse Cedex 4, 
France\\
E-mail : {\it sopik@irsamc.ups-tlse.fr ~\&~ 
clement.sire@irsamc.ups-tlse.fr ~\&~ 
chavanis@irsamc.ups-tlse.fr }
\vspace{0.5cm}
\end{center}


\begin{abstract}

We study the thermodynamical properties of a self-gravitating gas with
two or more types of particles. Using the method of linear series of
equilibria, we determine the structure and stability of statistical
equilibrium states in both microcanonical and canonical ensembles. We
show how the critical temperature (Jeans instability) and the critical
energy (Antonov instability) depend on the relative mass of the
particles and on the dimension of space. We then study the dynamical
evolution of a multi-components gas of self-gravitating Brownian
particles in the canonical ensemble. Self-similar solutions describing
the collapse below the critical temperature are obtained
analytically. We find particle segregation, with the scaling profile
of the slowest collapsing particles decaying with a non universal
exponent that we compute perturbatively in different limits. These
results are compared with numerical simulations of the two-species
Smoluchowski-Poisson system. Our model of self-attracting Brownian
particles also describes the chemotactic aggregation of a
multi-species system of bacteria in biology.

\end{abstract}

\maketitle


\begin{section}{Introduction}

In previous papers of this series \cite{prs}-\cite{chav}, we have
introduced a model of self-gravitating Brownian particles and we 
studied its equilibrium and collapse properties in the framework of
thermodynamics. In this model, the motion of the particles is
described by $N$ coupled stochastic equations (one for each particle)
involving a friction and a random force in addition to
self-gravity. The friction and the random force mimic the influence of
a thermal bath of non-gravitational origin imposing the
temperature. The temperature of the bath measures the strength of the
stochastic force. The self-gravitating Brownian gas model has a
conceptual interest in physics because it represents the canonical
counterpart of a Hamiltonian system of stars in Newtonian
interaction. Therefore, it can be used to test dynamically the
inequivalence of statistical ensembles which is generic for systems
with long-range interactions. Although most astrophysical systems are
described by the Newton equations (without dissipation), the
self-gravitating Brownian gas model could find applications for the
transport of dust particles in the solar nebula and the formation of
planetesimals by gravitational instability \cite{planetes}. In this
situation, the particles experience a drag force due to the friction
with the gas and a stochastic force due to turbulence. Furthermore,
self-gravity must be taken into account when the particles have grown
sufficiently by sticking processes and start to feel their mutual
attraction.  This would be just a first step because other ingredients
are required to improve the description of planetesimal formation. It
has also been shown in \cite{csr} that the process of violent relaxation 
for collisionless stellar systems exhibits similarities with the dynamics 
of a self-gravitating Brownian gas. In particular, the coarse-grained
distribution function $\overline{f}({\bf r},{\bf v},t)$ satisfies a
generalized Fokker-Planck equation, involving an effective diffusion
and an effective friction taking into account the peculiarities of the
collisionless evolution.

Our model of self-gravitating Brownian particles has also interest for
systems that are not necessarily related to astrophysics. For example,
in the physics of ultra cold gases, it has been shown recently that,
using a clever configuration of lasers beams, it is possible to
create an attractive $1/r$ interaction between atoms
\cite{kurihouches}. This leads to the fascinating possibility of
reproducing gravitational instabilities in the laboratory. In
particular, it is argued in \cite{od} that it should be possible to
observe the ``isothermal collapse'' \cite{c,fermions} of a Fermi gas cloud
in thermal equilibrium with a bosonic ``reservoir''. Since the system
is essentially {\it dissipative} a canonical description (fixed $T$)
is required and a plausible dynamical description of the system 
would be formed by the Fokker-Planck equation coupled to the
gravitational Poisson equation. In that case, the quantum nature of
the particles (fermions) is important, and generalized Fokker-Planck
equations, including the Pauli exclusion principle, must be considered as in
\cite{crrs}. On the other hand, as discussed in our previous papers, the
collapse of the self-gravitating Brownian gas is analogous to the
chemotactic aggregation of bacterial populations in biology. In
particular, the Smoluchowski-Poisson system which describes
self-gravitating Brownian particles in a strong friction limit is
isomorphic to a simplified version of the Keller-Segel model \cite{ks}
in biology, obtained in the limit of large diffusivity of the chemical
\cite{jager}. The Keller-Segel model is a standard model in
mathematical biology \cite{murray}. Due to this analogy, the results of
\cite{prs}-\cite{chav} have direct application for the chemotactic aggregation of bacterial populations.

For all these reasons, and also in its own right, the study of the
self-gravitating Brownian gas model \cite{prs} is clearly of interest
in physics.  So far, most works have focused on the case of a single
species of particles. In this paper, we extend these approaches to the
case of a multi-components system, with particular attention devoted
to the two-species model. In \sec{\ref{sec.analogy}}, we present the
basic equations describing a multi-components self-gravitating
Hamiltonian and Brownian system and show the analogies of the latter
with a multi-components chemotactic system. We use a mean-field
approach which is exact in a suitable thermodynamic limit $N_{\alpha}
\rightarrow +\infty$, keeping $\eta_{\alpha}=\beta G m^2_{\alpha}
N_{\alpha}/R^{d-2}$ constant for each species $\alpha$ (see Appendix
\ref{app.meanfield}).  In \sec{\ref{sec.static}}, we discuss the
statistical equilibrium states of a two-components self-gravitating
system in both microcanonical and canonical ensembles. Therefore, our
static study applies both to ordinary stellar systems (galaxies,
globular clusters,...) described in the microcanonical ensemble and
Brownian systems (or bacteria) described in the canonical ensemble. We
obtain the equilibrium density profiles and analyze their
thermodynamical stability by drawing the linear series of equilibria
(caloric curves) and using the turning point argument
\cite{katz}. We show how the critical temperature (Jeans instability
\cite{c}) and the critical energy (Antonov instability
\cite{antonov,lbw}) depend on the parameters $\mu= m_{1}/m_{2}$ and
$\chi=M_{1}/M_{2}$ where $m_{\alpha}$ is the individual mass of the
particles and $M_{\alpha}=N_{\alpha}m_{\alpha}$ the total mass of
species $\alpha$. In the microcanonical ensemble, we find that the
gravothermal catastrophe is advanced, i.e. it occurs sooner with
respect to the single-species case. In the canonical ensemble, the
isothermal collapse is advanced if we add heavy particles in the
system and delayed if we add lighter particles (keeping the total mass
fixed). Exact analytical expressions of the critical temperature of
collapse are given in dimension $d=2$. An approximate expression is
obtained for $d>2$ by using the Jeans swindle (see Appendix
\ref{app.jeans}). Our static study (\sec{\ref{sec.static}}) completes
previous investigations by Taff {\it et al.} \cite{taf} and De Vega \&
Siebert \cite{veg} in $d=3$ and Yawn \& Miller \cite{mil} in $d=1$.
In \sec{\ref{sec.collapse}}, we consider for the first time the
dynamics of the two-species self-gravitating Brownian gas in a strong
friction limit described by the Smoluchowski-Poisson system. We study
the collapse below $T_{c}$ by looking for self-similar
solutions. Depending on the values of $\mu$ and
$\zeta=\xi_{1}/\xi_{2}$, where $\xi_{\alpha}$ are the friction
coefficients, we show that the collapse of one species of particles
dominates the other. The invariant profile of the dominant species
scales as $\rho\sim r^{-2}$ as for the one-component gas
\cite{prs}. The collapse of the other particles is slaved to the
collapse of the dominant species. This decouples the equations of
motion and reduces the problem to the study of a single new dynamical
equation. We show that this equation possesses self-similar solutions
and that the scaling profile scales as $\rho\sim r^{-\alpha}$ where
$\alpha$ is a non-trivial exponent depending on $\mu$, $\zeta$ and
$d$, which leads to particle segregation. We determine this scaling
exponent perturbatively in a large dimension limit $d\rightarrow
+\infty$ on the one hand and for a weak asymmetry $\mu\rightarrow 1$
and $\zeta\rightarrow 1$ on the other hand. We also consider the
limits $\mu\rightarrow 0,+\infty$ or $\zeta\rightarrow
0,+\infty$. These perturbative analytical results are compared with
the exact results obtained numerically.

\end{section}

\begin{section}{Analogy between self-gravitating Brownian
particles and bacterial populations}
\label{sec.analogy}

\begin{subsection}{Self-gravitating Hamiltonian systems with
different species of particles}
\label{ssec.ham}

Let us consider a Hamiltonian system of $X$ species of particles
with mass $m_1,{...},m_X$ in a space of dimension $d$. Throughout
the paper, the particles of species $1$ are labeled from $i=1$ to
$g_1N$, the particles of species 2 from $(g_1 N + 1)$ to $g_2 N$,
and so on up to species $X$. The Latin letters $i$ will index the
$N$ particles and the Greek letters $\alpha$ will index the $X$
species. The particles interact via a long-range potential $U(
\textbf{r}_1,{...}, \textbf{r}_N) = \sum_{i<j} m_{i}m_{j}
u(\textbf{r}_i-\textbf{r}_j)$. In this paper, $u(\textbf{r}_i-
\textbf{r}_j) = - G/\lbrack {(d-2)|\textbf{r}_i-\textbf{r}_j|^{(d-2)}}
\rbrack$ denotes the gravitational potential of interaction in $d$ 
dimensions. This Hamiltonian system is completely defined by the 
equations of motion
\begin{eqnarray}
\label{eq.ham}
\begin{array}{rcl}
\displaystyle \frac{d \textbf{r}_i}{dt} &=& \textbf{v}_i \, ,
\vspace{6pt} \\
\displaystyle \frac{d \textbf{v}_i}{dt} &=& \displaystyle -
\frac{1}{m_i} \nabla_{i} U(\textbf{r}_1,{...},\textbf{r}_N).
\end{array}
\end{eqnarray}

In kinetic theory, the collisionless evolution of this system
is governed by the Vlasov-Poisson system, which is valid for
sufficiently ``short'' times. In fact, this regime can be extremely
long in practice since the relaxation time (Chandrasekhar's time)
increases almost linearly with the number of particles. The
collisional regime is usually described by the Landau-Poisson system
which governs the evolution of the distribution function $f({\bf r},
{\bf v},t)$ toward statistical equilibrium. For a multi-species
system in $d=3$, the Landau equation reads
\begin{eqnarray}
\label{eq.landau}
\frac{\partial f_{\alpha}}{\partial t} +
\textbf{v}\cdot\frac{\partial f_{\alpha}}{\partial \textbf{r}} +
\textbf{F}\cdot\frac{\partial f_{\alpha}}{\partial \textbf{v}}
= \frac{\partial}{\partial v^{\mu}} \sum_{\gamma = 1}^{X} \int
K^{\mu \nu} \left(m_{\gamma} f'_{\gamma}
\frac{\partial f_{\alpha}}{\partial v^{
\nu}} - m_{\alpha} f_{\alpha}\frac{
\partial f'_{\gamma}}{\partial v'^{\nu}}
\right)\,d{\bf v}', \nonumber \\
K^{\mu \nu} = 2 \pi G^2 \frac{1}{u} \ln\Lambda \left(\delta^{
\mu \nu} - \frac{u^{\mu}u^{\nu}}{u^2}\right),\qquad\qquad\qquad
\end{eqnarray}
where $\textbf{u} = \textbf{v} - \textbf{v}'$ is the relative velocity
of the particles involved in an encounter, $\ln \Lambda = \int_0^{+
\infty}\,dk/k$ is the Coulomb factor (which must be appropriately 
regularized) and  $\textbf{F} = -\nabla\Phi$ is the
gravitational force per unit of mass. We have also set
$f_{\alpha}'=f_{\alpha}({\bf r},{\bf v}',t)$ assuming that the 
encounters can be treated as local (see \cite{kandrup} for a critical 
discussion of this approximation). The gravitational potential 
$\Phi(\textbf{r},t)$ is determined by the Poisson equation
\begin{equation}
\label{eq.poissongen}
\Delta \Phi = S_d G \rho,
\end{equation}
with the total density $\rho= \sum_{\alpha=1}^X \rho_{\alpha}$, where
$\rho_{\alpha}({\bf r},t) = \int f_{\alpha}({\bf r},{\bf v},t)\,d
{\bf v}$ is the spatial density of species $\alpha$ and $f_{\alpha}({
\bf r},{\bf v},t)$ is their distribution function ($f_{\alpha}({\bf r},
{\bf v},t)\,d{\bf r}\,d{\bf v}$ gives the total mass of particles
of species $\alpha$ with position in (${\bf r}\,;\,{\bf r}+d{\bf r}$) 
and velocity in (${\bf v}\,;\,{\bf v}+d{\bf v}$) at time $t$). The total
distribution function is $f= \sum_{\alpha=1}^X f_{\alpha}$.

The Landau-Poisson system conserves the total mass
\begin{equation}
\label{eq.Malpha}
M_{\alpha} = \int \rho_{\alpha}\,d{\bf r}=N_{\alpha}m_{\alpha},
\end{equation}
of each species and the total energy
\begin{equation}
\label{eq.energy}
E=\frac{1}{2}\int f v^2\,d{\bf r}\,d{\bf v} +\frac{1}
{2}\int \rho \Phi\,d{\bf r}=K+W,
\end{equation}
where $K$ is the kinetic energy and $W$ is the potential energy.
Furthermore, the Landau-Poisson system satisfies a H-theorem
$\dot S \ge 0$ for the multi-components Boltzmann entropy
\begin{equation}
\label{eq.entropy}
S = - k_B \sum_{\alpha = 1}^{X} \int \frac{f_{\alpha}}{m_{\alpha}}
\ln\left(\frac{f_{\alpha}}{m_{\alpha}}\right) \,d{\bf r}\,d{\bf v}.
\end{equation}
At equilibrium, $\dot S=0$ implying that the current in the R.H.S
of \eqn{(\ref{eq.landau})} must vanish. The advective term in the
L.H.S of this equation must also vanish, independently. These two
conditions imply that the only stationary solution of the Landau
equation (\ref{eq.landau}) is the Maxwell-Boltzmann
distribution
\begin{equation}
\label{eq.falpha}
f_{\alpha}(\textbf{r},\textbf{v}) = A_{\alpha}\left(\frac{m_{\alpha}
\beta}{2\pi}\right)^{d/2}  e^{-\beta m_{\alpha}\left\lbrack\frac{v^{2}}{2}
+\Phi(\textbf{r})\right\rbrack},
\end{equation}
where the inverse temperature $\beta=1/k_BT$ appears as an integration
constant. Note that the advective term (Vlasov) is canceled out by
{\it any} distribution function $f_{\alpha}=f_{\alpha}(\epsilon)$
depending on the particle energy $\epsilon=\frac{v^{2}}{2}+\Phi({\bf
r})$ alone. The cancellation of the collision term singles out the
Boltzmann distribution among this infinite class of distributions. The
Maxwell-Boltzmann distribution \eqn{(\ref{eq.falpha})} represents the
statistical equilibrium state of the system in a mean-field
approximation. It can be obtained alternatively by maximizing the
entropy (\ref{eq.entropy}) at fixed energy and particle number (for
each species). The condition of thermodynamical stability in the
microcanonical ensemble ({\it maximum} of $S$ at fixed $E$,
$N_{\alpha}$) is equivalent to the linear dynamical stability with
respect to the Landau-Poisson system \cite{kinA}.

According to the theorem of equipartition of energy (which remains valid
here), the r.m.s. velocity of species $\alpha$ decreases with mass such
that
\begin{equation}
\label{eq.equi}
\langle v^{2}\rangle_{\alpha}=\frac{\int e^{-\beta m_{\alpha}\frac{v^{2}}
{2}}v^{2}d{\bf v}}{\int e^{-\beta m_{\alpha}\frac{v^{2}}{2}}d{\bf
v}}=d\frac{k_{B}T}{m_{\alpha}}.
\end{equation}
Therefore, heavy particles have less velocity dispersion to resist
gravitational attraction and will preferentially orbit in the inner
region of the system. This leads to mass segregation, but of a very
different nature from the dynamical segregation that we study in
Section 4. Defining the pressure by $p=\frac{1}{d}\int f v^{2}d{\bf
v}$, we get from \eqn{(\ref{eq.equi})} the local equation of state
\begin{equation}
\label{eq.eqstate}
p = \sum_{\alpha=1}^{X}\frac{\rho_{\alpha}}{m_{\alpha}}k_{B}T.
\end{equation}
The local mass density $\rho_{\alpha}$ of each species is obtained
directly from the integration of \eqn{(\ref{eq.falpha})} over
the velocities yielding
\begin{equation}
\label{eq.rhoalpha}
\rho_{\alpha}(\textbf{r}) = A_{\alpha} e^{- \beta m_{\alpha}
\Phi(\textbf{r})}.
\end{equation}
The gravitational field $\Phi({\bf r})$ is obtained 
self-consistently by substituting \eqn{(\ref{eq.rhoalpha})} in 
the Poisson equation (\ref{eq.poissongen}) and solving the 
resulting differential equation.

\end{subsection}

\begin{subsection}{Self-gravitating Brownian particles with different
species of particles}
\label{ssec.brown}

The Hamiltonian system of stars presented in \sec{\ref{ssec.ham}} is
associated to the microcanonical ensemble (fixed energy) in statistical
mechanics. We shall now introduce a model of particles in Newtonian
interaction associated with the canonical ensemble (fixed temperature).
Specifically, we consider a system of $N$ self-gravitating Brownian
particles belonging to $X$ different species. This is the generalization 
of the model introduced in \cite{prs}. This system is characterized by 
$N$ coupled stochastic equations
\begin{eqnarray}
\label{eq.stocha}
\begin{array}{rcl}
\displaystyle \frac{d \textbf{r}_i}{dt} &=& \textbf{v}_i,
\vspace{6pt} \\
\displaystyle \frac{d \textbf{v}_i}{dt} &=& \displaystyle - \xi_i
\textbf{v}_i - \frac{1}{m_i} \nabla_{i} U(\textbf{r}_1,{...},
\textbf{r}_N) + \sqrt{2 D_i} \textbf{R}_i(t),
\end{array}
\end{eqnarray}
where $\xi_i$ is the friction coefficient, $D_i$ is the diffusion
coefficient and $\textbf{R}_i(t)$ the stochastic force. In this paper
$\textbf{R}_i(t)$ is a white-noise satisfying the conditions $\left<
\textbf{R}_i(t)\right> = 0$ and $\left<R_{a,i}(t)R_{b,j}(t')\right> =
\delta_{ab} \delta_{ij}\delta(t-t')$ where $i,j$ refers to the particles
and $a,b$ to the space coordinates. The diffusion coefficient and the
friction coefficient are related to each other by the Einstein relation 
(see Appendix \ref{app.meanfield})
\begin{equation}
\label{eq.einstein}
D_{\alpha} = \frac{\xi_{\alpha} k_B T}{m_{\alpha}},
\end{equation}
where $T$ is the thermodynamical temperature. Therefore, the temperature
measures the strength of the stochastic force.

In the mean-field approximation, the evolution of the system is governed by
the multi-components  Kramers equation (see Appendix \ref{app.meanfield} for
details)
\begin{equation}
\label{eq.kramers}
\frac{\partial f_{\alpha}}{\partial t} + \textbf{v}\cdot\frac{\partial
f_{\alpha}}{\partial \textbf{r}} + \textbf{F}\cdot\frac{\partial f_{
\alpha}}{\partial \textbf{v}} = \frac{\partial}{\partial\textbf{v}}\cdot
\left(D_{\alpha}\frac{\partial f_{\alpha}}{\partial \textbf{v}} + \xi_{
\alpha}f_{\alpha}\textbf{v}\right),
\end{equation}
which must be coupled consistently with the Poisson equation, using
${\bf F}=-\nabla\Phi$. The Kramers-Poisson system conserves the total
mass of each species. Since the system is dissipative, the energy
(\ref{eq.energy}) is not conserved and the entropy (\ref{eq.entropy})
does not increase monotonically. However, introducing the free energy
\begin{equation}
\label{eq.freeenergy}
F[\lbrace f_{\alpha}\rbrace]= E[\lbrace f_{\alpha}\rbrace] - TS[\lbrace
f_{\alpha}\rbrace],
\end{equation}
the Kramers-Poisson system satisfies a sort of canonical H-theorem
$\dot F\le 0$. At equilibrium, $\dot F= 0$ implying that the diffusion
current in \eqn{(\ref{eq.kramers})} must vanish. The advective term
must also vanish. These two conditions lead to the Maxwell-Boltzmann
distribution (\ref{eq.falpha}) where $1/\beta$ is the
temperature of the bath. This distribution represents the statistical
equilibrium state of the system in a mean-field approximation. It can
be obtained alternatively by minimizing the free energy
(\ref{eq.freeenergy}) at fixed particle number (for each species). 
The condition of thermodynamical stability in the canonical
ensemble ({\it minimum} of $F$ at fixed $N_{\alpha}$) is equivalent to
the linear dynamical stability with respect to the Kramers-Poisson
system \cite{kinA}.

In order to simplify the problem further, we consider the strong friction
limit and let $\xi_i \rightarrow + \infty$ for each particle $i$. This
amounts to neglecting the inertia of the particles. Instead of 
\eqn{(\ref{eq.stocha})}, we obtain a simpler system of coupled stochastic
equations
\begin{equation}
\label{eq.stocha2}
\frac{d \textbf{r}_i}{dt} = - {\mu_i} \nabla_{i}U(\textbf{r}_1,{...},
\textbf{r}_N) + \sqrt{2 D'_i}\textbf{R}_i(t),
\end{equation}
where $\mu_{i}=1/m_{i}\xi_{i}$ is the mobility and
$D'_{i}=D_{i}/\xi_{i}^{2} =k_{B}T/m_{i}\xi_{i}$ is the diffusion
coefficient in physical space. The mean-field Fokker-Planck equation
obtained in this limit of strong friction is the Smoluchowski
equation, which can be written for each species
\begin{equation}
\label{eq.smogen}
\frac{\partial \rho_{\alpha}}{\partial t} = \frac{1}{\xi_{\alpha}} \nabla \cdot
\left(\frac{k_B T}{m_{\alpha}} \nabla \rho_{\alpha} + \rho_{\alpha} \nabla \Phi
\right).
\end{equation}
It has to be solved in conjunction with the Poisson equation
(\ref{eq.poissongen}). The passage from the Kramers to the
Smoluchowski equation can be made rigorous by using a
Chapman-Enskog expansion (see \cite{cll} for details and
generalizations). In the $\xi_{\alpha} \rightarrow +\infty$ 
limit, the distribution function can be written
\begin{equation}
\label{eq.falpha2}
f_{\alpha}(\textbf{r},\textbf{v},t) = \left(\frac{\beta m_{\alpha}}{2\pi}\right)
^{d/2}\rho_{\alpha} (\textbf{r},t) e^{-m_{\alpha} \beta \frac{v^2}{2}} + O\left(
\frac{1}{\xi_{\alpha}}\right),
\end{equation}
where $\rho_{\alpha}(\textbf{r},t)$ evolves according to
\eqn{(\ref{eq.smogen})}. Using \eqns{(\ref{eq.freeenergy})} and
(\ref{eq.falpha2}), it is possible to express the free energy as a
functional of the spatial density of each species in the form
\begin{equation}
\label{eq.freerho}
F[\lbrace \rho_\alpha\rbrace] = \frac{1}{2} \int\rho \Phi\,d{\bf r} + k_B T
\sum_{\alpha=1}^{X} \int \frac{\rho_{\alpha}}{m_{\alpha}}\ln\left(
\frac{\rho_{\alpha}}{m_{\alpha}}\right)\,d{\bf r},
\end{equation}
up to an irrelevant additive constant. The Smoluchowski-Poisson
system conserves the total mass of each species and decreases the free
energy $\dot F\le 0$. At equilibrium, the density is given by
\eqn{(\ref{eq.rhoalpha})}. The linearly dynamically stable steady
states minimize the free energy $F[\lbrace\rho_\alpha\rbrace]$ at
fixed mass (for each species) \cite{kinA}.

The Kramers and Smoluchowski equations can be written
\begin{equation}
\label{eq.kramers2}
\frac{\partial f_{\alpha}}{\partial t} + \textbf{v}\cdot\frac{\partial f_{\alpha}}
{\partial \textbf{r}} + \textbf{F}\cdot\frac{\partial f_{\alpha}}{\partial
\textbf{v}} = \frac{\partial}{\partial {\bf v}}\cdot \biggl \lbrack \xi_{\alpha}
f_{\alpha} \frac{\partial}{\partial {\bf v}}\biggl (\frac{\delta F}{\delta f_{
\alpha}}\biggr )\biggr\rbrack,
\end{equation}
\begin{equation}
\label{eq.smogen2}
\frac{\partial \rho_{\alpha}}{\partial t} = \frac{1}{\xi_{\alpha}} \nabla \cdot
\biggl ( \rho_{\alpha}\nabla \frac{\delta F}{\delta\rho_{\alpha}}\biggr ),
\end{equation}
where the free energy is respectively given by
\eqns{(\ref{eq.freeenergy})} and (\ref{eq.freerho}). They can also be
obtained from the linear thermodynamics of Onsager or by maximizing
the rate of free energy dissipation under appropriate constraints
\cite{fp}, which is the variational version of the linear
thermodynamics.

\end{subsection}

\begin{subsection}{Multi-components chemotactic systems}
\label{ssec.chemio}

In previous papers, see e.g. \cite{crrs}, we have shown that the equations
describing the dynamics of self-gravitating Brownian particles in a
strong friction limit were isomorphic to a simplified version of the
Keller-Segel model \cite{ks} describing the chemotactic aggregation of
bacterial populations.  We shall propose here a simple generalization
of this model to a multi-components system of bacteria and show the
relation with the multi-components Brownian model introduced
previously. Note that a more general multi-components chemotactic model
has been proposed recently by Wolansky \cite{w}.  We consider a system
of $X$ populations of bacteria with density $\rho_{\alpha}$, 
each species secreting a substance (chemical) with density $c_{\alpha}$. 
The bacteria diffuse with a diffusion coefficient $D_{\alpha}$
and they move along the (total) concentration of chemical $c=\sum_{\alpha}
c_{\alpha}$ as a result of a chemotactic attraction. The chemicals,
produced by the bacteria with a rate $a$, are degraded with a rate $b$.
They also diffuse with a diffusion coefficient $D'$.
The evolution of the system is described by the coupled differential
equations
\begin{eqnarray}
\label{eq.ksrho}
\frac{\partial \rho_{\alpha}}{\partial t} &=& D_{\alpha}\Delta\rho_{\alpha}
-\chi_{\alpha}\nabla (\rho_{\alpha}\nabla c),\\
\label{eq.ksc}
\frac{\partial c_{\alpha}}{\partial t} &=&  D'\Delta c_{\alpha}+a \rho_{
\alpha}-b c_{\alpha}.
\end{eqnarray}
Like in the one-species problem \cite{jager}, we shall consider a
regime of large diffusion of the chemicals so that we ignore the
temporal derivative in the second equation. We shall also take $b=0$,
assuming that there is no degradation of the chemicals. This reduces
the problem to the coupled system
\begin{eqnarray}
\label{eq.ksrho2}
\frac{\partial \rho_{\alpha}}{\partial t} &=& D_{\alpha}\Delta\rho_{\alpha}-\chi_{
\alpha}\nabla (\rho_{\alpha}\nabla c),\\
\label{eq.ksc2}
 \Delta c &=& -\lambda \rho.
\end{eqnarray}
These equations are isomorphic to the multi-components
Smoluchowski-Poisson system (\ref{eq.smogen})-(\ref{eq.poissongen})
provided that we make the identification
$D_{\alpha}=k_{B}T/\xi_{\alpha}m_{\alpha}$,
$\chi_{\alpha}=1/\xi_{\alpha}$, $c=-\Phi$ and $\lambda=S_{d}G$. {\it
Due to this analogy, the following results can be applied to the
chemotactic problem in biology by a proper reinterpretation of the
parameters}.

\end{subsection}

\end{section}

\begin{section}{Statistical equilibrium states of a multi-components system of
self-gravitating particles}
\label{sec.static}

\begin{subsection}{The thermodynamical potentials}
\label{ssec.mep}

At a fundamental level, the Boltzmann entropy is defined by
$S=k_{B}\ln W$ where $W$ is the number of microstates
(complexions) associated with a given macrostate. This number $W$
can be obtained by combinatorial analysis. In the continuum limit,
a macro-state is specified by the smooth distribution function
$f({\bf r},{\bf v})$ and the Boltzmann entropy takes the form of
\eqn{(\ref{eq.entropy})}. Therefore, if we assume that all
microstates are equiprobable for an isolated system at equilibrium
(microcanonical ensemble), the optimal distribution function
maximizes the Boltzmann entropy at fixed total energy and mass
(for each species). Introducing Lagrange multipliers and writing
the variational principle in the form
\begin{equation}
\label{eq.varentropy}
\delta S - \beta\delta E - \sum_{\alpha = 1}^{X}\lambda_\alpha \delta 
M_\alpha= 0,
\end{equation}
we obtain the Maxwell-Boltzmann distribution (\ref{eq.falpha}). It is
important to recall at that stage that the Boltzmann entropy has no
global maximum for self-gravitating systems. Hence, we have to confine
the system within a restricted region of space and look for {\it
local} entropy maxima. These metastable states are physically relevant
because their lifetime increases exponentially with the number of
particles \cite{metastable}.

On the other hand, if the system is in contact with a
heat bath fixing the temperature (canonical ensemble), the statistical
equilibrium state minimizes the free energy $F=E-TS$  at fixed mass 
(for each species). Introducing Lagrange multipliers and writing the
variational principle in the form
\begin{equation}
\label{eq.varfreeenergy}
\delta F  - \sum_{\alpha = 1}^{X}\lambda_\alpha \delta M_\alpha = 0,
\end{equation}
we obtain the Maxwell-Boltzmann distribution (\ref{eq.falpha}) as in
the microcanonical ensemble. What we have done essentially is a
Legendre transformation to pass from the entropy to the free energy, 
as the temperature is fixed instead of the energy. Here again, the 
system must be confined within a box and only {\it local} minima of 
free energy exist.

The statistical equilibrium distribution of particles is given by
\eqn{(\ref{eq.rhoalpha})} where the gravitational potential satisfies
the multi-species Boltzmann-Poisson equation
\begin{equation}
\label{eq.msbp}
\Delta\Phi=S_d G\sum_{\alpha=1}^{X}A_{\alpha}e^{-\beta m_{\alpha}\Phi}.
\end{equation}
In the microcanonical problem (Hamiltonian systems), the inverse
temperature must be related to the energy while in the canonical
problem (Brownian systems) it is imposed by the bath (and the
corresponding mean-field energy is interpreted as the averaged
energy). Then, we can plot the series of equilibria $\beta(E)$. The
stability of the system can be settled by the turning point argument
\cite{katz} as in the single-species case. Although the critical
points of constrained entropy and constrained free energy yield the
same density profiles, the stability limits (related to the sign of
the second order variations) will differ in microcanonical and
canonical ensembles. As these results on the inequivalence of
statistical ensembles have been extensively discussed in the
single-species case \cite{paddy,grand}, we shall not go into much
details here and rather focus on the new aspects brought by the
consideration of a distribution of mass among the particles. We also
recall that for systems with long-range interactions, the mean-field
description is exact (see Appendix \ref{app.meanfield}) so that our
thermodynamical approach is rigorous.

\end{subsection}

\begin{subsection}{The two-species Emden equation}
\label{ssec.emden}

From now on, we restrict ourselves to a system with only two species of
particles with mass $m_{1}$ and $m_{2}$. We assume that $m_{1}>m_{2}$
and set $\mu={m_{1}/ m_{2}}>1$. In order to determine the structure of isothermal
spheres, we introduce the function $\psi = m_2 \beta (\Phi- \Phi_0)$
where $\Phi_0$ is the gravitational potential at $r=0$. The density
profile of each species can then be written as
\begin{eqnarray}
\label{eq.rho}
\rho_{1} = \rho_{1}(0) e^{-\mu \psi} &,& \rho_{2} = \rho_{2}(0) e^{-\psi},
\end{eqnarray}
where $\rho_{1}(0)$ and $\rho_{2}(0)$ denote the central density. Restricting
ourselves to spherically symmetric solutions and introducing the notation
$\xi = (S_d G \beta m_2 \rho_2(0))^{1/2}r$, the Boltzmann-Poisson equation
(\ref{eq.msbp}) takes the dimensionless form
\begin{equation}
\label{eq.emden}
\frac{1}{\xi^{d-1}}\frac{d}{d\xi}\left(\xi^{d-1}\frac{d\psi}{d\xi}\right)
= e^{-\psi} + \lambda \mu e^{-\mu \psi},
\end{equation}
where $\lambda = n_{1}(0)/n_{2}(0)$ is the ratio of the central numerical
density $n_{\alpha}=\rho_{\alpha}/m_{\alpha}$ of the two species. Equation
(\ref{eq.emden}) represents the two-species Emden equation in $d$ dimensions.
It must be supplemented by the boundary conditions
\begin{equation}
\psi(0) = \psi'(0) = 0.
\end{equation}
The one-component case is recovered for $\lambda=0$.

The two-species Emden equation (\ref{eq.emden}) in dimension $d=3$ has
been studied by Taff {\it et al.} \cite{taf} who plotted the density
profiles and the caloric curves for different values of $\mu$. In
their work, the ratio $\lambda$ of central densities is maintained
fixed along the series of equilibria. We shall extend their study in a
space of dimension $d$ (with particular emphasis on the critical
dimension $d=2$) and consider the more physical (and more complicated)
case where the ratio $\chi={M_{1}/M_{2}}$ of the total mass of each
species (which are the conserved quantities) is kept fixed instead of
$\lambda$.  This makes possible to use the caloric curve $\beta(E)$ to
settle the thermodynamical stability of the system using the turning
point argument (this is not possible when $\chi$ varies along the
series of equilibria). Furthermore, we shall obtain analytical
expressions of the critical points (energy and temperature) as a
function of $\mu$ and $\chi$.

We shall first derive general properties of the differential equation
(\ref{eq.emden}). For $\xi \rightarrow 0$, an expansion of $\psi(\xi)$
in Taylor series yields
\begin{eqnarray}
\psi(\xi) &=& \frac{1 +\lambda \mu}{2d}\xi^2 - \frac{(1 +\lambda \mu)
(1 + \lambda \mu^2)}{8d(d+2)}\xi^4 \\
&& + \frac{1+ \lambda \mu}{48d^2(d+2)(d+4)} \lbrack d(1 + \lambda \mu^2)
^2+ (d+2)(1 + \lambda \mu)(1 + \lambda \mu^3)\rbrack\xi^6 + O(\xi^8).
\nonumber
\end{eqnarray}
To investigate the asymptotic behavior of $\psi(\xi)$ for $\xi \rightarrow
+\infty$, we first perform the transformation $t = \ln\xi$ and $z = - \psi
+ 2 \ln\xi$. In terms of $z$ and $t$, the two-species Emden equation
(\ref{eq.emden}) becomes
\begin{equation}
\label{eq.z}
\frac{d^2z}{dt^2} + (d-2) \frac{dz}{dt} = -\lambda \mu e^{\mu z}e^{-2(\mu-1)
t} - e^z + 2(d-2).
\end{equation}
For $\xi \rightarrow + \infty$, i.e. $t\rightarrow + \infty$, the
concentration of heavy particles, proportional to $e^{-\mu \psi}$,
goes to zero faster than the concentration of light particles,
proportional to $e^{-\psi}$, so the first term in the R.H.S. can be
neglected in a first approximation.  Then, \eqn{(\ref{eq.z})} 
reduces to the equation obtained for a single type of
particles. For $d>2$, it describes the damped motion of a fictitious
particle in a potential $V(z) = e^{z} - 2(d-2)z$ where $z$ plays the
role of position and $t$ the role of time. For $t \rightarrow + \infty$,
the system has reached its equilibrium position at $z_0 = \ln\lbrack2
(d-2)\rbrack$. Returning to initial variables, we find that $e^{-\psi}
\sim 2(d-2)/\xi^2$ for $\xi\rightarrow +\infty$. Since the two-species
Emden equation does not satisfy a homology theorem, this solution is
only valid asymptotically. It does not form a singular solution of 
\eqn{(\ref{eq.emden})} when $\lambda\neq 0$, contrary to the one-component
case \cite{sc}. We note that, for $d>2$, the total mass $M_{2}\sim \int_{0}^{+
\infty}\rho_{2}r^{d-1}dr$ of the lightest particles is infinite (as in
the single species case) since $\rho_{2} \sim r^{-2}$. However, since
$\rho_{1}\sim r^{-2\mu}$, the total mass of the heaviest particles is
finite if
\begin{equation}
\label{eq.mu32}
\mu>\mu_{3/2}=\frac{d}{2}.
\end{equation}

The next order correction to the asymptotic behavior of $\psi$ can be
obtained by setting $z = z_0 + z'$ with $z' \ll 1$ and keeping only
terms that are linear in $z'$. This yields
\begin{equation}
\label{eq.z'}
\frac{d^2z'}{dt^2} + (d-2) \frac{dz'}{dt} + 2(d-2)z' = -\lambda \mu 2^{
\mu}(d-2)^{\mu} e^{-2(\mu - 1)t}.
\end{equation}
This differential equation can be solved analytically. The discriminant
associated to the homogeneous equation exhibits two critical dimensions
$d = 2$ and $d = 10$ \cite{sc}. For $2<d<10$, we have for $\xi \rightarrow +
\infty$,
\begin{eqnarray}
\label{eq.epsi}
e^{-\psi} &=& \frac{2(d-2)}{\xi^2}  \left[1 + \frac{A}{\xi^{\frac{d-2}{2}}}
\cos\left(\frac{\sqrt{(d-2)(10-d)}}{2}\ln\xi + \delta\right)\right.\\
&& - \left.\frac{\lambda \mu 2^{\mu-1}(d-2)^{\mu}\xi^{-2(\mu - 1)}}{2(\mu-1)
^2-(d-2)(\mu-2)}\right], \nonumber
\end{eqnarray}
where $A$ and $\delta$ are integration constants. The density profile
(\ref{eq.epsi}) intersects the asymptotic solution $2(d-2)/\xi^2$ at
points that asymptotically increase geometrically in the ratio $1:e^{2\pi
/\sqrt{(d-2)(10-d)}}$. For
\begin{equation}
\label{eq.mu54}
\mu>\mu_{5/4}=\frac{d+2}{4},
\end{equation}
the last term in \eqn{(\ref{eq.epsi})} can be neglected for sufficiently
large $\xi$ and there is an infinite number of intersections. For $\mu<
\mu_{5/4}$, there is only a finite number of intersections. For $d>10$,
we have for $\xi \rightarrow +\infty$,
\begin{eqnarray}
\label{eq.epsi2}
e^{-\psi} &=& \frac{2(d-2)}{\xi^2} \left[1 + \frac{1}{\xi^{\frac{d-2}{2}}}
\left(A \xi^{\frac{\sqrt{(d-2)(d-10)}}{2}} + \frac{B}{\xi^{\frac{\sqrt{(d-2)
(d-10)}}{2}}}\right)\right. \\
&& - \left.\frac{\lambda \mu 2^{\mu-1}(d-2)^{\mu}\xi^{-2(\mu-1)}}{2(\mu-1)^2
-(d-2)(\mu-2)}\right].\nonumber
\end{eqnarray}
There is no intersection with the asymptotic solution. For $d<2$, the density
profile of the lightest particles decreases as $e^{-\psi}\sim C e^{-A_{d}
\xi^{2-d}}$  and for $d=2$ as $e^{-\psi}\sim A \xi^{-\delta}$. The normalized density
profiles are plotted in \fig{\ref{fig.rho13d}} in $d=1$ and $d=3$. The case $d=2$
is postponed to \sec{\ref{ssec.2d}}.

\begin{figure}
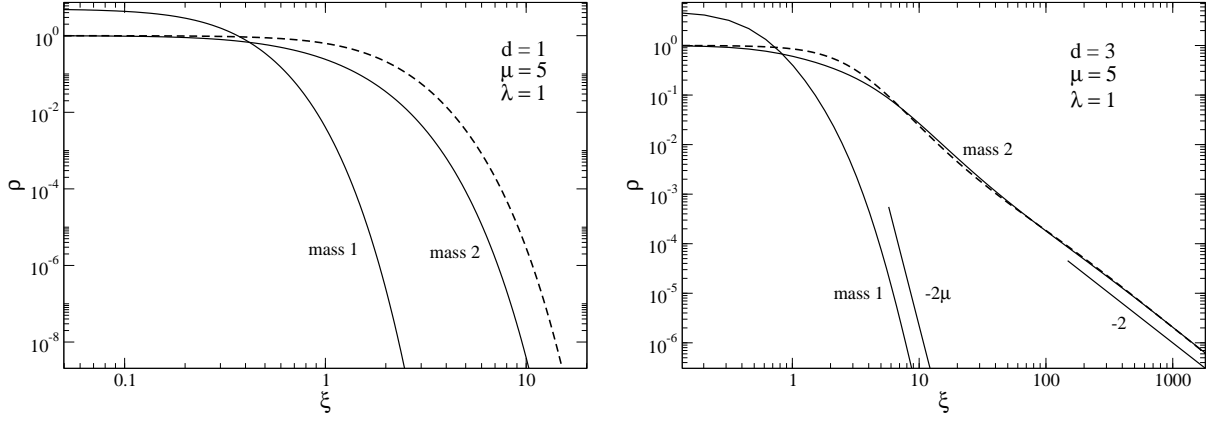

\vskip0.5cm
\centerline{
\psfig{figure=rho.1d.l1.eps,angle=0,height=5.5cm}
\hspace{6pt}
\psfig{figure=rho.l1.eps,angle=0,height=5.5cm}}
\caption{\label{fig.rho13d}The dimensionless density profiles $\tilde{\rho}_1(\xi)
= \lambda \mu e^{-\mu\psi}$ and $\tilde{\rho_2}(\xi) = e^{-\psi}$ for $\mu = 5$
and for $\lambda = 1$ in $d=1$ (\fig{\ref{fig.rho13d}}.a) and $d=3$ 
(\fig{\ref{fig.rho13d}}.b). The dashed line represents the density of the one component 
system. }
\end{figure}

\end{subsection}

\begin{subsection}{The Milne variables}
\label{ssec.milne}

Since the multi-species Emden equation does not satisfy a homology theorem,
it cannot be transformed into a first order differential equation as in the
one-species case. However, the use of the Milne variables is still useful
to analyze the phase portrait of the equation. In the general case, they are
defined by
\begin{eqnarray}
\label{eq.varuv}
u = \frac{d\ln M(r)}{d\ln r} = \xi \frac{\lambda \mu e^{-\mu \psi}
+ e^{-\psi}}{\psi'} &,& v = - \frac{d\ln p(r)}{d\ln r} = \xi \psi'
\left(\frac{\lambda \mu e^{-\mu \psi} + e^{-\psi}}{\lambda e^{-\mu
\psi} + e^{-\psi}}\right),
\end{eqnarray}
where we used the integrated density $M(r)=\int_{0}^{r} \rho S_{d}
r^{d-1}\,dr$ and the total pressure $p(r)=(\rho_1/m_1 +\rho_2/m_2)
k_B T$.  Taking the logarithmic derivatives of $u$ and $v$ with respect
to $\xi$ and introducing the notation $w = (\lambda e^{- \mu\psi} +
e^{-\psi})(\lambda \mu^2 e^{- \mu \psi} + e^{-\psi})/(\lambda\mu e^{-\mu
\psi} + e^{-\psi})^2$, we get
\begin{eqnarray}
\label{eq.dudxi}
\frac{1}{u}\frac{du}{d\xi} &=& \frac{1}{\xi}(d - u - v w),\\
\label{eq.dvdxi}
\frac{1}{v}\frac{dv}{d\xi} &=& \frac{1}{\xi}[u + v(1-w) - (d-2)].
\end{eqnarray}
The single species case is recovered for $\lambda=0$ and $w=1$. Taking
the ratio of the above equations, we obtain
\begin{equation}
\label{eq.dvdu}
\frac{u}{v}\frac{dv}{du} = \frac{u + v(1-w) - (d-2)}{d - u - v w}.
\end{equation}

The solution curve in the $(u,v)$ plane is plotted in \fig{\ref{fig.uv13d}}.
The curve is parameterized by $\xi$. It starts from the point $(u,v)=(d,0)$
with a slope
\begin{equation}
\label{eq.dvdu0}
\left(\frac{dv}{du}\right)_0 = -\frac{d+2}{d} \frac{(1+\lambda \mu)^3}
{(1+\lambda)[1+\lambda \mu(\mu(1+\lambda \mu) + 1)]},
\end{equation}
corresponding to $\xi=0$.  For $d>2$ and $\xi\rightarrow +\infty$, the
curve converges to the limit point $(d-2,2)$ which corresponds to
the asymptotic behavior $e^{-\psi} \sim 2(d-2)/\xi^2$. Contrary to the
single-species case, the ($u,v$) curve can make loops before spiraling
around the limit point. These loops are a signature of a 
multi-components system : using \eqn{(\ref{eq.dvdu})}, the points of
horizontal and vertical tangent are defined respectively by
$u+v(1-w)=d-2$ and $u+vw=d$. Due to the term $w$, new solutions of
these equations arise with respect to the single-species case and they
create loops. For $d>10$, the $(u,v)$ curve reaches the limit point
without spiraling but still makes loops for the reason previously
mentioned.  For $d<2$, the curve tends monotonically to $(0,+\infty)$
for $\xi \rightarrow +\infty$ as in the single species case. The
two-dimensional case is discussed in
\sec{\ref{ssec.2d}}.

\begin{figure}
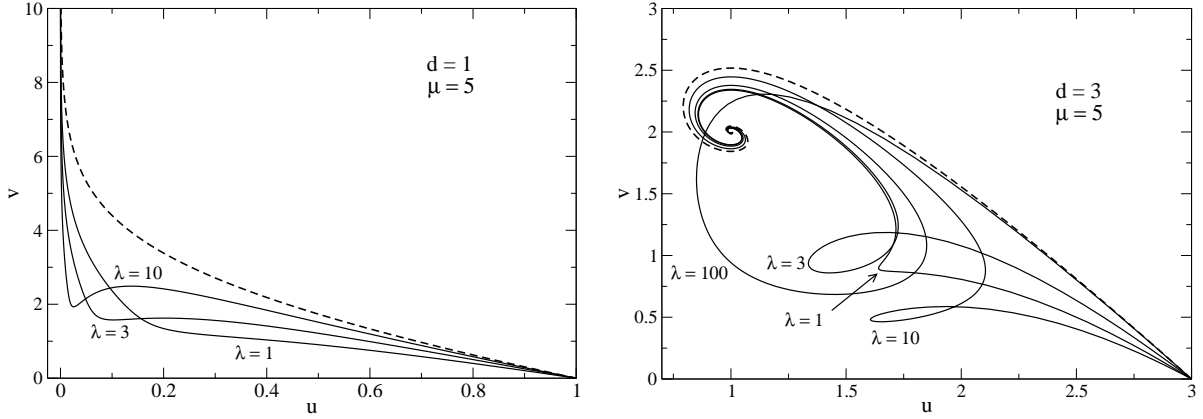

\vskip0.5cm
\centerline{
\psfig{figure=uv.1d.mu5.eps,angle=0,height=5.5cm}
\hspace{6pt}
\psfig{figure=uv.mu5.eps,angle=0,height=5.5cm}}
\caption{\label{fig.uv13d}The solution of the two-species Emden equation in
the ($u,v$) plane for $d=1$ (left, \fig{\ref{fig.uv13d}}.a) and in $d = 3$ 
(right, \fig{\ref{fig.uv13d}}.b). The single-species case is represented by 
the dashed line.}
\end{figure}

\end{subsection}

\begin{subsection}{The thermodynamical parameters}
\label{ssec.thermo}

As indicated previously, isothermal self-gravitating systems have
infinite mass. We shall overcome this problem by confining the system
within a spherical box of radius $R$ (Antonov model). Physically, the
box delimits the region of space where thermodynamical arguments
can be applied. In the biological problem (chemotaxis), the box
represents the natural boundary of the domain in which the bacteria
live. For bounded isothermal systems, the solution of 
\eqn{(\ref{eq.emden})} is terminated by the box at a normalized radius
given by $\alpha =(S_d G m_2 \beta\rho_2(0))^{1/2} R$. We shall now
determine the temperature and the energy corresponding to the
configuration indexed by $\alpha$.

Using the Poisson equation (\ref{eq.poissongen}), we write the Gauss
theorem
\begin{equation}
GM =G \int  \rho \,d{\bf r} = S_d G \int_0^R \rho r^{d-1} \, dr =
\int_0^R\frac{d}{dr} \left(r^{d-1}\frac{d\Phi}{dr}\right) \, dr =
\left(r^{d-1}\frac{d\Phi}{dr}\right)_{r=R}.
\end{equation}
Introducing  the dimensionless variables defined previously, we
find that the normalized inverse temperature is given by
\begin{equation}
\label{eq.eta}
\eta \equiv \frac{\beta G M m_2 }{R^{d-2}} = \alpha \psi'(\alpha).
\end{equation}

The calculation of the energy $E = K + W$ is a little more intricate.
The kinetic energy is given by
\begin{equation}
\label{eq.K}
K = \frac{d}{2} \left(N_1 + N_2\right) k_B T = \frac{d}{2}\left(\frac{
M_1}{m_1} + \frac{M_2}{m_2}\right) k_B T = \frac{d}{2} \frac{M_2}{m_2}
\left(\frac{\chi}{\mu} + 1\right)k_B T.
\end{equation}
Using $M = M_2 (\chi + 1)$ and \eqn{(\ref{eq.eta})}, the normalized kinetic
energy can be written
\begin{equation}
\label{eq.Knorm}
- \frac{KR^{d-2}}{GM^2} = - \frac{d}{2\alpha \psi'(\alpha)} \frac{\chi
+\mu}{\mu(\chi + 1)}.
\end{equation}
For $d \ne 2$, the expression of the potential energy can be deduced
from the Virial theorem
\begin{equation}
\label{eq.virial}
2K+(d-2)W=dV_{d}R^{d}p(R),
\end{equation}
where $V_d=S_d/d$ is the volume of a $d$-dimensional 
sphere with unit radius (and $S_d$ is the surface of a
$d$-dimensional unit sphere) \cite{lang}. Using 
$p(R)=({\rho_{1}(R)/m_{1}}+{\rho_{2}(R)/m_{2}})k_{B}T$ and the
expressions (\ref{eq.rho}) of the density, we directly obtain
\begin{equation}
\label{eq.Wnorm}
-\frac{W R^{d-2}}{G M^2} = - \frac{1}{(d - 2)\psi'^2(\alpha)}
\left(\lambda e^{-\mu \psi(\alpha)} + e^{-\psi(\alpha)}\right)
+ \frac{d}{(d-2)\alpha \psi'(\alpha)} \frac{{\chi}+
{\mu}}{\mu(\chi + 1)}.
\end{equation}
Adding \eqns{(\ref{eq.Knorm})} and (\ref{eq.Wnorm}), we find that the
total normalized energy is
\begin{equation}
\label{eq.L1}
\Lambda \equiv -\frac{ER^{d-2}}{GM^{2}}=\frac{d}{2\alpha \psi'(\alpha)}
\biggl (\frac{4 - d}{d - 2}\biggr )\frac{{\chi}+{\mu}}{\mu(\chi + 1)}
- \frac{1}{(d - 2)\psi'^2(\alpha)} \left(\lambda e^{-\mu \psi(\alpha)}
+ e^{-\psi(\alpha)}\right).
\end{equation}
Note that an alternative expression of the potential energy, valid
also for $d=2$, can be obtained along the following lines. Starting
from the expression
\begin{equation}
\label{eq.W}
W = \frac{1}{2} \int \rho \Phi\,d{\bf r},
\end{equation}
and introducing the dimensionless variables defined previously, we get
\begin{equation}
\label{eq.Wnorm2}
- \frac{WR^{d-2}}{GM^2} = - \frac{1}{2 \alpha^d \psi'^2(\alpha)}
\int_0^\alpha \left(\lambda \mu e^{-\mu \psi} + e^{- \psi}\right)(\psi
+ \psi_0) \xi^{d-1}\,d\xi,
\end{equation}
where $\psi_0 = m_2 \beta \Phi(0)$ represents the normalized central
gravitational potential. It is determined by the relation $\psi(\alpha)
= m_2 \beta \left(\Phi(R) - \Phi(0)\right)$ with $\Phi(R)=-{G M}/\lbrack
{(d-2) R^{d-2}}\rbrack$ for $d \ne 2$. This yields
\begin{equation}
\psi_0 =-\biggl ( \frac{\alpha \psi'(\alpha)}{d-2} + \psi(\alpha)\biggr ).
\end{equation}
Equation (\ref{eq.Wnorm2}) remains valid for $d=2$ but in that case, 
$\Phi(R)=0$ so that $\psi_{0}=-\psi(\alpha)$. The corresponding expression 
of the total normalized energy is now
\begin{equation}
\label{eq.L2}
\Lambda = -\frac{d}{2\alpha \psi'(\alpha)} \frac{{\chi}+{\mu}}{\mu(\chi + 1)}
- \frac{1}{2 \alpha^{d} \psi'^2(\alpha)} \int_0^\alpha \left(\lambda \mu e^{
-\mu \psi} + e^{- \psi}\right)(\psi + \psi_0) \xi^{d-1}\,d\xi.
\end{equation}

Equations (\ref{eq.eta}) and (\ref{eq.L1}) define a series of equilibria 
$\beta(E)$ parameterized by the value of the normalized radius $\alpha$, or
equivalently by the density contrast ${\cal R}\equiv\rho_{2}(0)/\rho_{2}(R)
=e^{\psi(\alpha)}$. Along this series of equilibria, we can either fix the
ratio of central densities $\lambda$ or the ratio of total mass $\chi$.
These two parameters are related to each other by
\begin{equation}
\label{eq.chi}
\chi \equiv  \frac{M_1}{M_2} = \frac{\lambda \mu \int_0^\alpha \xi^{d-1}
e^{-\mu \psi}\,d\xi} {\int_0^\alpha \xi^{d-1} e^{-\psi}\,d\xi}.
\end{equation}
In the framework of statistical mechanics, it is more relevant to fix
$\chi$ along the series of equilibria since the total mass of each
species is a conserved quantity. Furthermore, it is only under this
condition that the turning point argument can be used to settle the
stability of the system. Therefore, in the foregoing equations,
$\lambda$ must be viewed as an implicit function of $\alpha$ given by
\begin{equation}
\label{eq.l}
\lambda(\alpha) = \frac{\chi}{\mu} \frac{\int_0^\alpha \xi^{d-1}
e^{-\psi}\,d\xi}{\int_0^\alpha \xi^{d-1} e^{-\mu \psi}\,d\xi}.
\end{equation}
Then, for given $\alpha$, the two-species Emden equation (\ref{eq.emden})
must be solved by an iterative procedure in order to satisfy the constraint
(\ref{eq.l}).

\begin{figure}
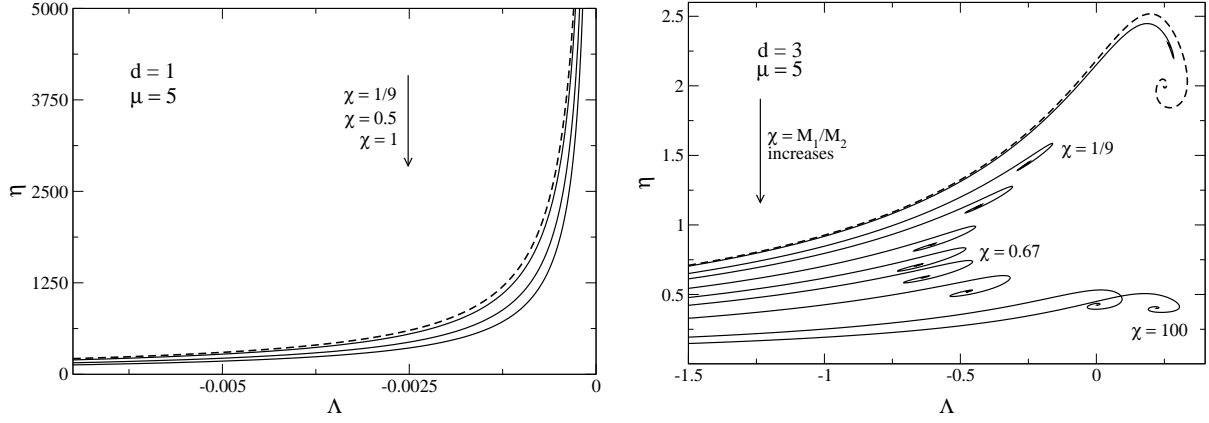

\vskip0.5cm
\centerline{
\psfig{figure=etalambda.1d.mu5.mfix.eps,angle=0,height=5.5cm}
\hspace{6pt}
\psfig{figure=etalambda.mu5.mfix.eps,angle=0,height=5.5cm}}
\caption{\label{fig.etalambda13d}Series of equilibria (caloric curves) for a
two-components isothermal gas in $d = 1$ (left, \fig{\ref{fig.etalambda13d}}.a) 
and in $d = 3$ (right, \fig{\ref{fig.etalambda13d}}.b). We plot the inverse 
normalized temperature $\eta = G M m_2 \beta/R^{d-2}$ as a function of the 
normalized energy $\Lambda = - E R^{d-2}/G M^2$ for several values of the 
total mass ratio $\chi=M_1/M_2$. The dashed curve represents the one 
component case, i.e. $\chi = 0$.}
\end{figure}

Figure \ref{fig.etalambda13d} displays an ensemble of caloric curves
in $d = 1$ and in $d = 3$ for different values of $\chi$ (at fixed
$\mu$). In $d=1$, the curves are monotonic and the system is always
stable. In $d=3$, the curves present turning points at which a mode of
stability is lost depending on the ensemble considered (a vertical
tangent corresponds to a loss of microcanonical stability and a
horizontal tangent to a loss of canonical stability). These results
have been discussed in detail for the one-species case, see
e.g. \cite{c}, and will not be repeated here.  We shall just discuss
how the critical points (beyond which no equilibrium state exists)
depend on the relative mass of the particles.  First, consider the
canonical ensemble in which the control parameter is the normalized
inverse temperature $\eta$. For $\eta>\eta_{c}(\mu,\chi)$, the system
undergoes an ``isothermal collapse''.  For $\chi = 0$ we obviously
recover the value $\eta_c \simeq 2.52$ of the single species case. As
the mass ratio $\chi$ increases (at fixed total mass $M$ and $\mu>1$),
$\eta_c$ decreases ($T_c$ increases) up to $\eta_{c}=2.52/\mu$
obtained for $\chi \rightarrow +\infty$. In the microcanonical
ensemble, the control parameter is the normalized energy
$\Lambda$. For $\Lambda>\Lambda_{c} (\mu,\chi)$, the system undergoes
a ``gravothermal catastrophe''. For $\chi=0$ and $\chi \rightarrow
+\infty$, we recover the single-species value $\Lambda_c\approx
0.335$. Between these two extreme values, $\Lambda_c$ passes by a
minimum $(\Lambda_{c})_{min}(\mu)$.  These results are illustrated in
\fig{\ref{fig.eclc3d}} where $\eta_c$ and $\Lambda_c$ are plotted as a
function of $\chi$ for a given value of $\mu$. The value of the
minimum of the normalized energy $(\Lambda_{c})_{min}(\mu)$ seems to
behave linearly with $\mu$ (except for $\mu\rightarrow 1$) as
illustrated in \fig{\ref{fig.Lcmin}}. If we take the particles of mass
$m_{2}$ as a reference, we conclude that the onset of isothermal
collapse in the canonical ensemble is advanced when heavier particles
$m_{1} >m_{2}$ are added to the system (keeping the total mass $M$
fixed). It is delayed if lighter masses ($\mu<1$) are added. On the
other hand, the onset of the gravothermal catastrophe in the
microcanonical ensemble is always advanced in a multi-species system
(with respect to the single species case), whatever the mass of the
particles added (keeping the total mass $M$ fixed).

\begin{figure}
\vskip0.5cm
\centerline{
\psfig{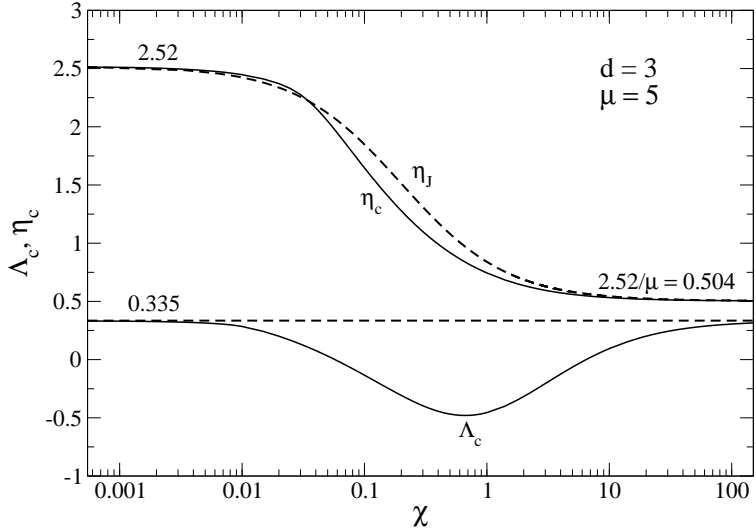}}
\caption{\label{fig.eclc3d}Evolution of the critical normalized inverse
temperature $\eta_c$ (Jeans temperature) and the critical normalized
energy $\Lambda_c$ (Antonov energy) as a function of $\chi$ for $d=3$
and $\mu=5$. We plot with a dashed-dotted line the critical
temperature $\eta_J$ obtained by using the Jeans swindle 
(see Appendix \ref{app.jeans} for details). Note this ``na\"{\i}ve'' 
prediction provides a reasonable fit of the exact critical
temperature $\eta_c$.}
\end{figure}

\begin{figure}
\vskip0.5cm
\centerline{
\psfig{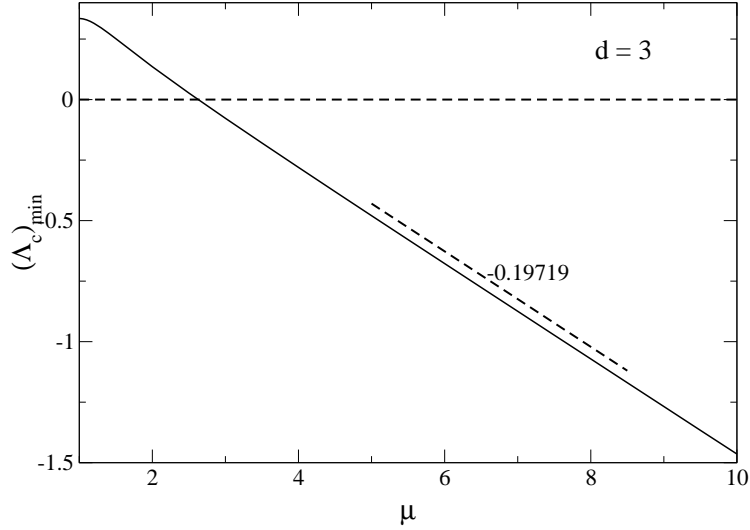}}
\caption{\label{fig.Lcmin} Evolution of the minimum Antonov energy
$(\Lambda_c)_{min}$ as a function of $\mu$ for $d=3$. In the range
considered, it decreases approximately linearly as $-0.19719
\mu$. Note that the minimum Antonov energy becomes positive for
$\mu\simeq 2.627$ and $\chi\simeq 0.68$. Furthermore, the value of
$\chi$ for which $\Lambda_c$ is minimum is always close to $0.7$
(except for $\mu\rightarrow 1$). This is probably related to the fact
that $(\Lambda_c)_{min}(\mu)$ is almost linear in this range.  }
\end{figure}

Some analytical results can be obtained for $\alpha\rightarrow+\infty$
and $d>2$. This corresponds to the configurations located near the
limit point in \fig{\ref{fig.etalambda13d}}.b, at the center of the
spiral. In that case, it will be shown a posteriori that
$\lambda(\alpha)$ diverges for a fixed $\chi$. Accordingly, we can
neglect the term $e^{-\psi}$ in the Emden equation (\ref{eq.emden})
which reduces to
\begin{equation}
\label{eq.modifemden1b}
\frac{1}{\xi^{d-1}}\frac{d}{d \xi} \left(\xi^{d-1} \frac{d \psi}{d
\xi}\right) = \lambda \mu e^{-\mu \psi}.
\end{equation}
This approximation is valid for $\xi < \xi'$, where $\xi'$ is such
that $\lambda \mu e^{-\mu \psi (\xi')} \sim e^{-\psi(\xi')}$. If we
introduce a new potential $\Theta$ depending on $\zeta \equiv
\sqrt{\lambda} \mu \xi$ through the defining relation
\begin{equation}
\label{eq.newemdenvar}
\psi(\xi) = \frac{1}{\mu} \Theta\left(\zeta\right) = \frac{1}{\mu}
\Theta\left(\sqrt{\lambda} \mu \xi\right),
\end{equation}
then \eqn{(\ref{eq.modifemden1b})} takes the form of the ordinary Emden
equation
\begin{equation}
\label{eq.emden1b}
\frac{1}{\zeta^{d-1}}\frac{d}{d \zeta} \left(\zeta^{d-1} \frac{d\Theta}
{d \zeta}\right) = e^{- \Theta}.
\end{equation}
Using the behavior $\Theta \sim 2 \ln\zeta - \ln\lbrack 2(d-2)\rbrack$
for large $\zeta$, we obtain the following behavior of $\psi(\xi)$ in
the range $1\ll \xi<\xi'$:
\begin{eqnarray}
\label{eq.largexipsi}
e^{-\psi}\sim \frac{\lbrack 2(d-2)\rbrack^{1/\mu}}{(\sqrt{\lambda}\mu
\xi)^{2/\mu}}.
\end{eqnarray}
We shall find a posteriori that $\xi'\sim \alpha$ so that the range of
validity of this behavior is huge in the limit $\alpha\rightarrow
+\infty$. This scaling in $\xi^{-2/\mu}$ contrasts from the scaling in
$\xi^{-2}$ obtained in the limit $\xi\rightarrow +\infty$ for fixed
$\lambda$. The validity of \eqn{(\ref{eq.largexipsi})} is confirmed in
\fig{\ref{fig.rho3d}} where we plot the normalized density profile of
a bounded isothermal system for a large value of $\alpha$.

\begin{figure}
\vskip0.5cm
\centerline{
\psfig{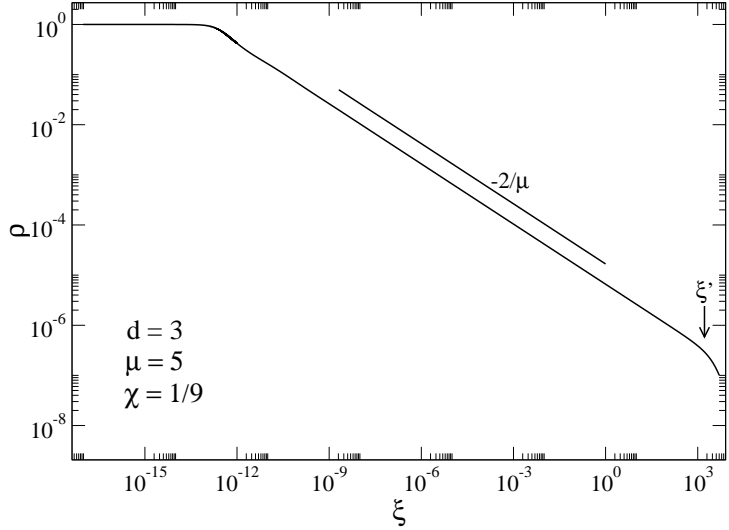}}
\caption{\label{fig.rho3d}Dimensionless density profile $\tilde{\rho}_2
(\xi) = e^{-\psi}$ of the lightest particles enclosed within a box for
$d=3$, $\mu = 5$, $\chi = 1/9$ and $\alpha=5000$. In the limit $\alpha
\rightarrow +\infty$, the profile decays as $\xi^{-2/\mu}$ for $1\ll
\xi<\xi' \sim \alpha$. This can be contrasted to the $\xi^{-2}$ decay 
for $\xi \rightarrow +\infty$ in an open system with fixed $\lambda$ (see 
\fig{\ref{fig.rho13d}}). }
\end{figure}

Using \eqns{(\ref{eq.l})} and (\ref{eq.largexipsi}), we can investigate
the asymptotic behavior of $\lambda(\alpha)$ for $\alpha\rightarrow
+\infty$. We have to estimate the two integrals
\begin{eqnarray}
\label{eq.I1alpha}
I_1(\alpha) &=& \int_0^{\alpha} \xi^{d-1} e^{-\mu \psi}\,d\xi,\\
\label{eq.I2alpha}
I_2(\alpha) &=& \int_0^{\alpha} \xi^{d-1} e^{-\psi}\,d\xi,
\end{eqnarray}
for large values of $\alpha$. We check that for $d>2$ and $\mu>1$, the
integrals (extended to $+\infty$) do not converge. Therefore, both terms 
in the decomposition
\begin{equation}
\label{eq.I2alphabis}
I_2(\alpha) = \int_0^{\xi'} \xi^{d-1} e^{-\psi}\,d\xi+\int_{\xi'}^{\alpha}
\xi^{d-1} e^{-\psi}\,d\xi
\end{equation}
behave as a power law with the same exponent (the second integral is
not negligible with respect to the first). We can obtain the
asymptotic behavior of the first integral by using the analytical
expression (\ref{eq.largexipsi}) of $\psi$ in the range $1\ll\xi<\xi'$. 
However, since we do not know the expression of $\psi$ for 
$\xi'<\xi<\alpha$, we cannot compute the second integral. Thus we can
get the exponent of the power law divergence of $I_{i}(\alpha)$ but
not the prefactor.

\begin{figure}
\vskip0.5cm
\centerline{ 
\psfig{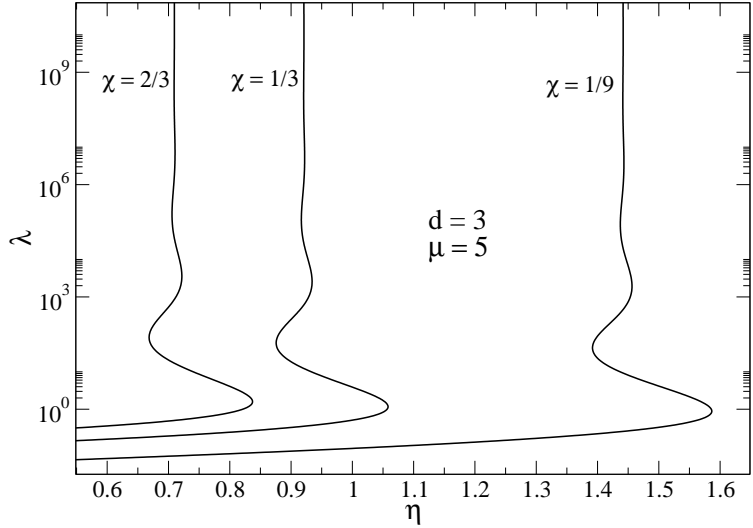}}
\caption{\label{fig.leta3d}The $\lambda(\eta)$ curves for several
values of $\chi = M_1/M_2$ for $d = 3$. These curves are
parameterized by $\alpha$.}
\end{figure}

Evaluating \eqns{(\ref{eq.I1alpha})} and (\ref{eq.I2alpha}) with 
\eqn{(\ref{eq.largexipsi})}, we obtain
\begin{eqnarray}
\label{eq.Ialpha1}
I_1(\alpha) \sim K_{1}\frac{\alpha^{d-2}}{\lambda(\alpha)} &,& I_2(\alpha)
\sim K_{2} \frac{\alpha^{d -{2}/{\mu}}}{\lambda(\alpha)^{{1}/{\mu}}},
\end{eqnarray}
where, for the reasons explained previously, the prefactors are not
known. The asymptotic behavior of $\lambda(\alpha)$ is now obtained
by substituting Eq. (\ref{eq.Ialpha1}) in \eqn{(\ref{eq.l})}. This yields
\begin{eqnarray}
\label{eq.lalpha1}
\lambda(\alpha) \sim K \alpha^{2(\mu-1)}.
\end{eqnarray}
As $\mu > 1$, the numerical density ratio $\lambda(\alpha)$ always
diverges for $\alpha\rightarrow +\infty$ and $d > 2$. This justifies
our initial assumptions. If we now insert \eqn{(\ref{eq.lalpha1})} in
(\ref{eq.Ialpha1}), we find that $I_2(\alpha)$ diverges as 
$I_2(\alpha) \sim \alpha^{d-2}$. On the other hand, $I_{1}(\alpha)$ 
behaves as $I_{1}(\alpha) \sim \alpha^{d-2\mu}$ which diverges for 
$\mu<\mu_{3/2}=d/2$ and tends to zero for $\mu>\mu_{3/2}$. In 
\fig{\ref{fig.leta3d}} we plot the ratio of central numerical densities 
$\lambda$ as a function of $\eta$ for different values of $\chi$.

\end{subsection}

\begin{subsection}{The two-dimensional case}
\label{ssec.2d}

The dimension $d=2$ is a critical dimension for self-gravitating
systems \cite{sc}. It is also the relevant dimension for the biological 
problem of chemotaxis, since bacterial colonies usually live on a plate.
Therefore, the dimension $d=2$ requires a particular attention. The
two-dimensional Emden equation (\ref{eq.emden}) reads
\begin{equation}
\label{eq.emden2d}
\frac{1}{\xi}\frac{d}{d \xi}\left(\xi \frac{d \psi}{d \xi}\right)
 = \lambda \mu e^{- \mu
\psi} + e^{-\psi}.
\end{equation}
The density profile behaves asymptotically as $e^{-\psi}\sim A
\xi^{-\delta}$ where $A$ and $\delta$ are constants. For $\lambda=0$
(single-species case), \eqn{(\ref{eq.emden2d})} can be solved
analytically and we get the exponent $\delta=4$.  For other values
of $\lambda$, we have $\delta\neq 4$. Some density profiles
are plotted in \fig{\ref{fig.rho2d}}. The phase portrait of the Emden
equation (\ref{eq.emden}) in the Milne plane is shown in 
\fig{\ref{fig.uv2d}}. On the other hand, the thermodynamical parameters
$\eta$ and $\Lambda$ are given by
\begin{equation}
\label{eq.eta2d}
\eta = G M m_2 \beta = \alpha \psi'(\alpha),
\end{equation}
\begin{equation}
\label{eq.L2d}
\Lambda = -\frac{E}{G M^2} = - \frac{1}{\alpha \psi'(\alpha)} 
\frac{{\chi}/{\mu} + 1}{\chi + 1} - \frac{1}{2 \alpha^2\psi'^2(\alpha)} 
\int_0^\alpha \left(\lambda \mu e^{-\mu \psi} + e^{-\psi}\right)
(\psi -\psi(\alpha)) \xi\,d\xi.
\end{equation}
Note that the normalized temperature and the normalized energy do not
depend on $R$. This is a consequence of the logarithmic form of the
gravitational potential in $d = 2$. An ensemble of caloric curves are
plotted in \fig{\ref{fig.etalambda.2d.mfixe}}. As in the previous
section, the value of $\chi$ is fixed along a series of equilibria so
that $\lambda(\alpha)$ is determined by an iterative procedure. In
continuity with the single species case, the caloric curves form a
plateau for $\Lambda\rightarrow +\infty$. Thus, there exists a
critical inverse temperature $\eta_{c}(\chi,\mu)$ above which no
equilibrium state is possible in the canonical ensemble (by contrast,
there is no critical energy in $d=2$ in the microcanonical
ensemble). The critical temperature $\eta_{c}(\chi,\mu)$ has two
expressions depending on whether $\mu>2$ or $\mu<2$ as we now show.

\begin{figure}
\vskip0.5cm
\centerline{
\psfig{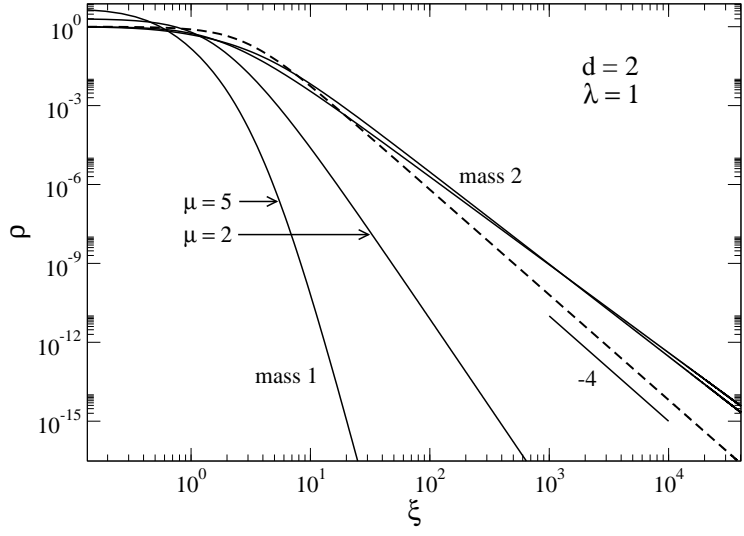}}
\caption{\label{fig.rho2d} Dimensionless density profiles
$\tilde{\rho}_1(\xi) = \lambda \mu e^{-\mu\psi}$ and
$\tilde{\rho_2}(\xi) = e^{-\psi}$ in dimension $d = 2$ for 
$\lambda = 1$ and for $\mu = 2$ and $\mu = 5$. For comparison, 
the dashed line represents the density of the singe-component 
system with slope $-4$. The slope of the profile $e^{-\psi} \sim
\xi^{-\delta}$ depends on $\lambda$ and $\mu$.}
\end{figure}

\begin{figure}
\vskip0.5cm
\centerline{
\psfig{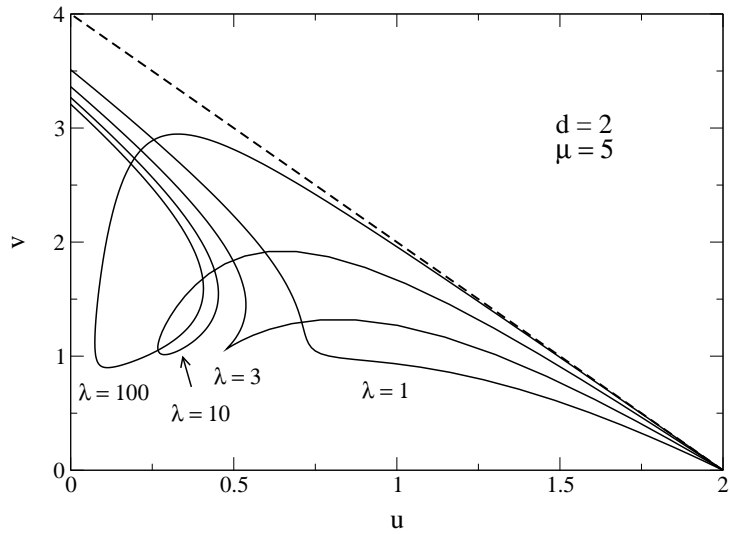}}
\caption{\label{fig.uv2d}The solution of \eqn{(\ref{eq.emden})} in
the ($u,v$) plane, where $u$ and $v$ are defined by 
\eqn{(\ref{eq.varuv})} for the two-components system at 
$d = 2$. The single-species case is represented by the dashed 
line. All the curves start at $(u,v) = (d,0)$ with the initial 
slope given by \eqn{(\ref{eq.dvdu})}. For $\xi\rightarrow +\infty$, 
they tend to the terminal point $(0,\delta)$.}
\end{figure}

\begin{figure}
\vskip0.5cm
\centerline{
\psfig{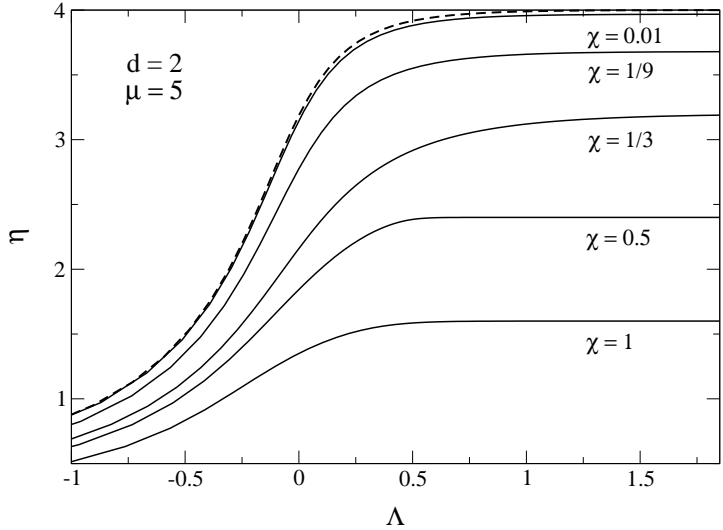}}
\caption{\label{fig.etalambda.2d.mfixe}An ensemble of caloric
curves for different values of  $\chi = M_1/M_2$ in the 
two-dimensional case. The dashed curve represents the 
one-component case.}
\end{figure}

We first consider the situation in which, at $T=T_{c}$, the two
profiles form a Dirac peak at $r=0$. The parameters corresponding to
this situation will be found a posteriori. In this case, the critical
temperature can be obtained from the Virial theorem as in the
single-species problem \cite{lang}. We start from the general 
relation valid in $d=2$ :
\begin{equation}
\label{eq.virialeq}
2K - \frac{GM^2}{2} = 2p(R)V,
\end{equation}
with $V=\pi R^{2}$. If the densities are concentrated in a Dirac peak
at $r=0$, the total pressure at the edge of the box vanishes.
Equations (\ref{eq.K}) and (\ref{eq.virialeq}) directly lead to the
result
\begin{equation}
\label{tcc}
k_B T_c  = \frac{GM^{2}}{4N}.
\end{equation}
The critical normalized temperature is
\begin{equation}
\label{eq.etac1}
\eta_c = 4\frac{{\chi}+{\mu}}
{\mu(\chi + 1)}.
\end{equation}
For $\chi=0$, we recover the critical inverse temperature $\eta_{c}=4$
obtained for the single-species case \cite{sc}.

It will be shown that the above regime, called regime (I), corresponds
to the case where the function $\lambda(\alpha)$ converges for $\alpha
\rightarrow +\infty$. We now consider the regime (II) where 
$\lambda(\alpha)$ diverges so that \eqn{(\ref{eq.emden2d})} reduces to 
\eqn{(\ref{eq.modifemden1b})} with $d=2$. With the change of variables
of \eqn{(\ref{eq.newemdenvar})}, we obtain the classical Emden equation
(\ref{eq.emden1b}). In $d=2$, it can be solved analytically and, returning 
to original variables, we get
\begin{equation}
\label{eq.psi2d1}
\psi(\xi) = \frac{2}{\mu} \ln\left(1 + \frac{\lambda \mu^2}{8}\xi^2
\right).
\end{equation}
This analytical expression provides a good approximation of the solution 
for all values of $\xi < \xi'$ where $\xi'$ is such that
\begin{equation}
\label{eq.xipr}
\int_{0}^{\xi'} e^{-\psi(\xi)}\xi\,d\xi\sim \lambda\mu  \int_{0}^{\xi'} 
e^{-\mu\psi(\xi)}\xi\,d\xi.
\end{equation}
Substituting \eqn{(\ref{eq.psi2d1})} into \eqn{(\ref{eq.xipr})} and 
using \eqn(\ref{eq.lalpha2}), we find that $\xi'\sim\alpha$ so 
the range of validity of the analytical expression is huge, as 
checked numerically. The density profile of the lightest particles 
decreases as $e^{-\psi}\sim \xi^{-4/\mu}$ and the density profile 
of the heaviest particles decreases as $e^{-\mu\psi}\sim \xi^{-4}$. 
These heaviest particles form a Dirac peak as $\alpha\rightarrow +\infty$. Furthermore, the density $\lambda \mu e^{-\mu\psi}$ of species $1$ becomes smaller 
than the density $e^{-\psi}$ of species $2$ for $\xi>\xi''\sim 
\alpha^{-(\mu-2)^{2}/(4(\mu-1))}\rightarrow 0$. In order to determine 
the asymptotic behavior of $\lambda(\alpha)$ for $\alpha\rightarrow +\infty$, 
we have to estimate the integrals $I_1(\alpha)$ and $I_2(\alpha)$ defined by
\eqns{(\ref{eq.I1alpha})} and (\ref{eq.I2alpha}). We will consistently
show that the assumptions made in regime (II) are only valid for $\mu
> 2$. Then, it is easy to show that the integral $I_{1}$ extended to
$+\infty$ is convergent while the integral $I_{2}$ is divergent. 
Therefore, by using the profile (\ref{eq.psi2d1}) to
evaluate (\ref{eq.I1alpha}) and (\ref{eq.I2alpha}), we obtain the
exact asymptotic expression of $I_{1}$ while we only obtain the
correct exponent of $\alpha$ in $I_{2}$ but not the prefactor. Indeed,
in the calculation of $I_{1}$ the second integral in the decomposition
(\ref{eq.I2alphabis}) is negligible while in the calculation of
$I_{2}$ it is of the same order as the first.  Then, we obtain
\begin{eqnarray}
\label{eq.I1I2}
I_1(\alpha) \sim\frac{4}{\lambda \mu^2} &,& I_2(\alpha) \sim K_{2} 
\lambda^{-2/\mu} \alpha^{{2(\mu - 2)}/{\mu}},
\end{eqnarray}
and, using \eqn{(\ref{eq.l})}, we get
\begin{eqnarray}
\label{eq.lalpha2}
\lambda(\alpha) \sim \left(\frac{\chi K_{2}\mu}{4}\right)^{\frac{\mu}{2}}
\alpha^{\mu - 2},
\end{eqnarray}
for $\alpha \rightarrow + \infty$. The assumption that $\lambda(\alpha)$ 
diverges is only consistent with $\mu > 2$. Note that from 
\eqn{(\ref{eq.lalpha2})} and the value of $I_{1}$, we obtain
the exact result $I_{2}\rightarrow 4/\chi\mu$. We also note that 
$I_{1}\rightarrow 0$ for $\alpha\rightarrow +\infty$.

\begin{figure}
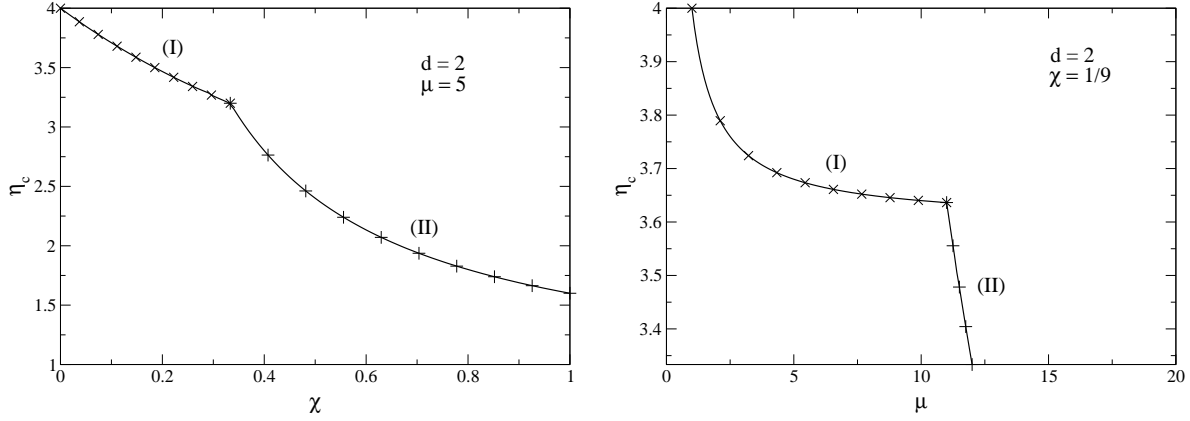

\vskip0.5cm
\centerline{
\psfig{figure=etac.mu5.eps,angle=0,height=5.5cm}
\hspace{6pt}
\psfig{figure=etac.chi0.111.eps,angle=0,height=5.5cm}}
\caption{\label{fig.etac}The critical temperature $\eta_c$ is
plotted versus $\chi$ (\fig{\ref{fig.etac}}.a) and $\mu$
(\fig{\ref{fig.etac}}.b). The solid lines represent the numerical
results. They are in excellent agreement with the theoretical results
(\ref{eq.etac2}) and (\ref{eq.etac1}) in the two regimes. The 
transition between these regimes appears when $\chi =
\chi_* = 1/(\mu - 2)$ in \fig{\ref{fig.etac}}.a and when $
\mu = \mu_* = 2 + 1/\chi$ in \fig{\ref{fig.etac}}.b.}
\end{figure}

\begin{figure}
\vskip0.5cm
\centerline{
\psfig{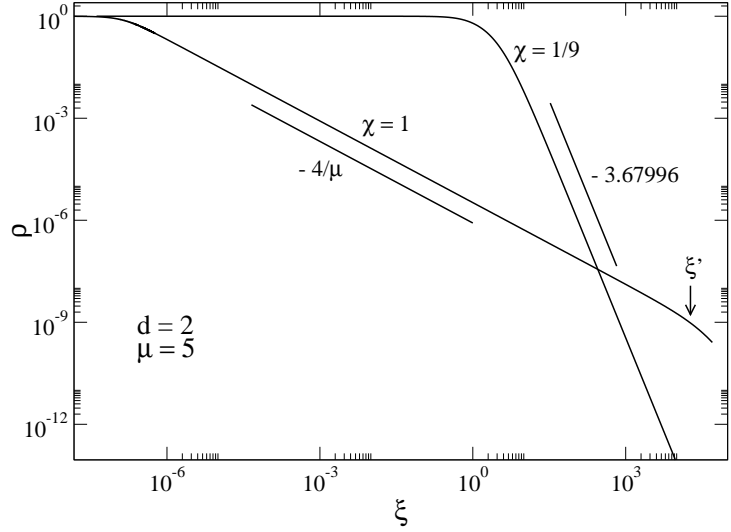}}
\caption{\label{fig.rho2d.mfixe}The normalized density profile $
\tilde{\rho}_2 = e^{-\psi}$ is plotted for $\mu = 5>2$ and for two
different values of $\chi$ situated from both sides of the critical
value $\chi_* = 1/(\mu-2) = 1/3$. For $\chi=1/9 < \chi_*$ (regime I),
the asymptotic slope of the profile is $-\eta_c=-3.6799$ given by
\eqn{(\ref{eq.etac1})}.  For $\chi=1 > \chi_*$ (regime II), the
asymptotic slope in the range $1\ll\xi\ll \xi'\sim\alpha$ is $-4/\mu = 
-0.8$, see \eqn{(\ref{eq.psi2d1})}. The final asymptotic slope in the 
range $\xi'< \xi<\alpha$ is $-\eta_c=-8/5$ given by \eqn{(\ref{eq.etac2})}.}
\end{figure}

We are now able to obtain the critical inverse temperature $\eta_c$ in
regime (II). In this regime, the heavy particles form a Dirac peak at
$r=0$ for $T=T_{c}$ while the light particles extend in the whole
box. Since their density is non-zero on the edge of the box, we cannot
use the reasoning valid in regime (I). However, the normalized temperature 
(\ref{eq.eta2d}) can be written in the form
\begin{equation}
\eta = \beta G M m_2 = 2\pi \beta G m_2  \left(1 + \frac{1}{\chi}\right)
\int_{0}^{R} \rho_1(r)\, rd{r}= \lambda \mu \left(1 +
\frac{1}{\chi}\right) \int_0^{\alpha} e^{- \mu \psi} \xi\,d\xi,
\end{equation}
where the last integral is precisely $I_{1}(\alpha)$. Using \eqn{(\ref{eq.I1I2})} 
for $\alpha\rightarrow +\infty$, we get
\begin{equation}
\label{eq.etac2}
\eta_c = \frac{4}{\mu} \left(1 + \frac{1}{\chi}\right).
\end{equation}
We note the fortunate cancellation of $\lambda$ which allows one to
obtain this exact result without detailed knowledge of $K_{2}$. We now
have two different expressions of the critical normalized inverse
temperature $\eta_c$, resp. \eqns{(\ref{eq.etac1})} and (\ref{eq.etac2}). 
We find that the crossover between the two regimes is obtained for
\begin{eqnarray}
\label{eq.critmassratio}
\chi_* = \frac{1}{\mu - 2} &,& \mu_{*}=2+\frac{1}{\chi}.
\end{eqnarray}
Applying the Virial theorem in regime (II), we can determine the exact
expression of the normalized density of species $2$ on the box. Indeed, 
using \eqns{(\ref{eq.virialeq})} and (\ref{eq.etac2}), we obtain      
\begin{eqnarray}
\label{eq.bord}
\alpha^{2}e^{-\psi(\alpha)}\rightarrow \frac{8\lbrack \chi(\mu-2)-1\rbrack}
{\mu^{2}\chi^{2}},
\end{eqnarray}
for $\alpha\rightarrow +\infty$. This implies another necessary
condition to be satisfied in regime (II), namely $\chi>\chi_{*}$.

In conclusion, regime (I) corresponds to $\mu \le 2$ and ($\mu>2$ and
$\chi<\chi_{*}$) ; in that case, $\lambda(\alpha)$ converges and the
critical temperature is given by \eqn{(\ref{eq.etac1})}. 
Regime (II) corresponds to ($\mu > 2$ and $\chi>\chi_{*}$) ; in that 
case, $\lambda(\alpha)$ diverges and the critical temperature is given by
\eqn{(\ref{eq.etac2})}. Equivalently, for a given $\chi$, if
$\mu<\mu_*$ (regime I) the critical temperature is given by
\eqn{(\ref{eq.etac1})} while if $\mu>\mu_*$ (regime II) it is given by
\eqn{(\ref{eq.etac2})}. Figure \ref{fig.etac} clearly exhibits the
cross-over of these two different regimes. The theoretical predictions
(\ref{eq.etac1}), (\ref{eq.etac2}) of the critical temperature are
perfectly consistent with the numerical results. We plot the
normalized density $\tilde{\rho}_2 = e^{-\psi}$ in
\fig{\ref{fig.rho2d.mfixe}}. In the regime (I) where $\lambda(\alpha)$ 
converges, we have the asymptotic behavior $e^{-\psi}\sim A\xi^{-\delta}$ 
for $\xi\rightarrow +\infty$. Using the fact that $\xi\psi'(\xi)
\rightarrow \eta_{c}$ for $\xi\rightarrow +\infty$, we find that 
$\delta=\eta_{c}$ where $\eta_{c}$ is explicitly given by
\eqn{(\ref{eq.etac1})}. In particular, $\delta=\eta_{c}=4$ for the
single species case.  In the regime (II) where $\lambda(\alpha)$
diverges, we have the asymptotic behavior $e^{-\psi}\sim
A\xi^{-4/\mu}$ for $1\ll \xi<\xi'$ and $\alpha\rightarrow
+\infty$. The final asymptotic slope in the range $\xi'< \xi<\alpha$
is $-\eta_c$ given by \eqn{(\ref{eq.etac2})}. The numerical results
are fully compatible with these predicted values.

We now address the determination of the critical temperature in $d=2$
for a system with more than two types of particles. If all the species
collapse on a Dirac peak at $T=T_{c}$ as in regime (I) discussed
previously, the critical temperature is again given by \eqn{(\ref{tcc})}
with $N=\sum_{\alpha}N_{\alpha}$ and $M=\sum_{\alpha}N_{\alpha}m_{\alpha}$. 
The critical normalized inverse temperature $\eta=\beta GMm_{X}$ is
\begin{eqnarray}
\label{seo}
\eta_{c}=\frac{4Nm_{X}}{M}.
\end{eqnarray}
Alternatively, we can consider the case where $\lambda_{1}(\alpha)$
diverges for $\alpha\rightarrow +\infty$, as in regime (II) discussed
previously. To apply the same approximations as before, we need to
have $\lambda_{1}\gg \lambda_{i}$ for $i=2,...,X$ where $\lambda_{i}=
n_{i}(0)/n_{X}(0)$. Repeating the steps described previously in this 
more general situation, we find that $\lambda_{1}\rightarrow +\infty$ 
if $\mu_{1}>2$ and $\lambda_{1}\gg\lambda_{i}$ if $\mu_{1}>2\mu_{i}$. 
We shall assume that these conditions are fulfilled (i.e. $m_{1}>2m_{2}$). 
In that case $\lambda_{1}\sim \alpha^{(\mu_{1}-2)}$ and we find by an approach
similar to that described previously that
\begin{eqnarray}
\label{seog}
\eta_{c}=\frac{4Mm_{X}}{M_{1}m_{1}}.
\end{eqnarray}

\end{subsection}

\end{section}

\begin{section}{Collapse of a multi-components system}
\label{sec.collapse}

\begin{subsection}{Self-similar solutions of the two-components
Smolu\-chow\-ski-Pois\-son system}
\label{ssec.general}

We now consider the dynamics of a system of self-gravitating Brownian
particles. We restrict ourselves to the case of only two types of mass
$m_{1}$ and $m_{2}$, as we shall see that the general case of a {\it
discrete} spectrum of particles is a simple generalization of this
problem. We also restrict our analysis to a spatial dimension $d>2$. The
dimension $d=2$ is critical and deserves a particular treatment (see
\cite{sc} for the single species case). As in our previous works, we
consider a limit of strong friction $\xi\rightarrow +\infty$ so that
the dynamical equations reduce to the two-species Smoluchowski-Poisson
system (\ref{eq.smogen}). We also restrict ourselves to spherically
symmetric solutions. By introducing dimensionless variables, we can
set $k_B=G=R=M=m_{1}=\xi_{1}=1$ without loss of generality. Then, the
problem depends only on the asymmetry parameters
$\mu=m_{1}/m_{2}=1/m_{2}$ and $\zeta=\xi_{1}/\xi_{2}=1/ \xi_{2}$ and
on the temperature $T=1/(\eta\mu)$. With these conventions, the
dynamical equations can be written
\begin{eqnarray}
\label{eq.smo}
\begin{array}{rcl}
\displaystyle \frac{\partial \rho_1}{\partial t} &=& \displaystyle
\nabla \cdot (T \nabla \rho_1 + \rho_1 \nabla \Phi),
\vspace{6pt} \\
\displaystyle \frac{\partial \rho_2}{\partial t} &=& \displaystyle
\zeta \nabla \cdot (T\mu \nabla \rho_2 + \rho_2 \nabla\Phi),
\end{array}
\end{eqnarray}
\begin{eqnarray}
\label{eq.poisson}
&\Delta \Phi = S_d \rho.&
\end{eqnarray}
We shall impose a vanishing flux across the surface of the confining
sphere. Therefore, the boundary conditions are
\begin{equation}
\begin{array}{rcl}
\displaystyle \frac{\partial \Phi(0,t)}{\partial r} = 0, &&
\displaystyle \Phi(1,t) = \frac{1}{2-d},
\vspace{6pt} \\
\displaystyle T \frac{\partial \rho_{1}}{\partial r}(1,t)+\rho_1(1,t)=0
,&& \displaystyle T\mu \frac{\partial\rho_{2}}{\partial r}(1,t)+
\rho_2(1,t) = 0.
\end{array}
\end{equation}
Using the Gauss theorem, we can rewrite the Smoluchowski-Poisson system
(\ref{eq.smo})-(\ref{eq.poisson}) in the form of two integrodifferential
equations
\begin{eqnarray}
\label{eq.intdiff}
\begin{array}{rcl}
\displaystyle \frac{\partial \rho_1}{\partial t} &=&
\displaystyle \frac{1}{r^{d-1}} \frac{\partial}{\partial r}
\left[r^{d-1}\left(T\frac{\partial \rho_1}{\partial r} +
\frac{\rho_1}{r^{d-1}}\int_0^r S_d \rho(r') r'^{d-1} \, dr'
\right)\right],%
\vspace{6pt} \\
\displaystyle \frac{\partial \rho_2}{\partial t} &=&
\displaystyle \frac{\zeta}{r^{d-1}} \frac{\partial}{\partial
r} \left[r^{d-1}\left(T\mu \frac{\partial \rho_2}{\partial r}
+ \frac{\rho_2}{r^{d-1}}\int_0^r S_d \rho(r') r'^{d-1} \, dr'
\right)\right].
\end{array}
\end{eqnarray}
The Smoluchowski-Poisson system (\ref{eq.intdiff}) is also
equivalent to a set of two coupled differential equations
\begin{eqnarray}
\label{eq.intmass}
\begin{array}{rcl}
\displaystyle \frac{\partial M_1}{\partial t} &=& \displaystyle
T \left(\frac{\partial^2 M_1}{\partial r^2} +  \frac{1-d}{r}
\frac{\partial M_1}{\partial r}\right) + \frac{M_1 + M_2}{r^{
d-1}}\frac{\partial M_1}{\partial r},
\vspace{6pt} \\
\displaystyle \frac{\partial M_2}{\partial t} &=& \displaystyle
\zeta \left\lbrack T \mu \left(\frac{\partial^2 M_2}{\partial r^2}
+ \frac{1-d}{r}\frac{\partial M_2}{\partial r}\right) + \frac{M_1 
+ M_2}{r^{d-1}}\frac{\partial M_2}{\partial r}\right\rbrack,
\end{array}
\end{eqnarray}
for the quantities
\begin{equation}
M_{\alpha}(r,t) = \int_0^r  \rho_{\alpha}(r',t) S_d r'^{d-1}\,dr',
\end{equation}
which give the mass of species $\alpha=1,2$ within a sphere of
radius $r$. In terms of these variables, the boundary conditions
take the form
\begin{equation}
M_{\alpha}(0,t) =0 \qquad ; \qquad {M_1(1,t)} = \frac{\chi}{1+\chi}, 
\qquad M_2(1,t) = \frac{1}{1+\chi} .
\end{equation}
Note that we shall restrict ourselves to the pre-collapse regime, so
that we do not consider the possibility that a Dirac peak forms at
$r=0$. A Dirac peak forms in the post-collapse regime for $d>2$ and in
$d=2$ (see \cite{sic,sc} in the single species case). It will be more
convenient to work in terms of the functions $s_{\alpha}(r,t) =
M_{\alpha}(r,t)/r^d$ which have the dimension of a density. They
satisfy
\begin {equation}
\label{eq.s}
\begin{array}{rcl}
\displaystyle \frac{\partial s_1(r,t)}{\partial t} &=&
\displaystyle T \left(\frac{\partial^2 s_1}{\partial r^2} +
\frac{d+1}{r} \frac{\partial s_1}{\partial r}\right) + (s_1 +
s_2) \left(r\frac{\partial s_1}{\partial r} + d s_1\right),
\vspace{6pt} \\
\displaystyle \frac{\partial s_2(r,t)}{\partial t} &=&
\displaystyle\zeta \left\lbrack T \mu \left(\frac{\partial^2 s_2}
{\partial r^2} + \frac{d+1}{r}\frac{\partial s_2}{\partial r}
\right) + (s_1 + s_2) \left(r\frac{\partial s_2}{\partial r}
+ d s_2\right)\right\rbrack.
\end{array}
\end{equation}
We look for self-similar solutions of the form
\begin{eqnarray}
\label{eq.ansatz}
s_{1}(r,t) = \rho_{0}(t)S_{1}\left(\frac{r}{r_0(t)}\right) &,&
s_{2}(r,t) = \rho_{0}^{\alpha/2}(t) S_{2}\left(\frac{r} {r_0(t)}
\right),
\end{eqnarray}
where $\rho_{0}(t)$ represents the typical central density of species
$1$ and $r_{0}(t)$ is the typical core radius (of the two species)
defined by
\begin{equation}
\label{eq.rho0r0}
\rho_0 r_0^2 = T.
\end{equation}
On physical grounds, we expect that the total density should scale as
in the single-species case because, on a coarse-grained scale the fine
structure of the mass distribution should not matter (except for a
continuous spectrum of mass going from $[0,+\infty[$ with peculiar
behavior at the extremes, which is not the case here). Therefore,
either the two profiles scale the same manner or one dominates the 
other. Now, by solving numerically the scaling equation coming from 
\eqns{(\ref{eq.s})}-(\ref{eq.ansatz}), we have found that the problem 
does not admit any physical solution with $\alpha=2$. Hence one species 
will dominate the other. We define species $1$ as the one that 
dominates the dynamics. This choice imposes $\alpha<2$ for the
other species. We will give later the conditions on $\mu$ and 
$\zeta$ for which this requirement is satisfied. Inserting 
\eqn{(\ref{eq.ansatz})} in \eqn{(\ref{eq.s})} and using the notation 
$x = r/r_0(t)$, the equation for $s_{1}(r,t)$ is transformed into
\begin{eqnarray}
\label{eq.scaling1b}
\frac{d\rho_0}{dt}S_1(x) - x\frac{\rho_0}{r_0}\frac{dr_0}{dt}S'_1(x)
&=& T \left(\frac{\rho_0}{r_0^2}S''_1(x) + \frac{d + 1}{x} \frac{\rho_0}
{r_0^2}S'_1(x)\right) \\
&& + \left(\rho_0 S_1(x) + \rho_0^{\alpha/2}S_{2}(x)\right)\left(
x\rho_0 S'_1(x) + d \rho_0 S_1(x)\right). \nonumber
\end{eqnarray}
For sufficiently high densities, we can neglect the sub-dominant
term $\rho_0^{\alpha/2} S_{2}(x)$ in the above equation. Then,
\eqn{(\ref{eq.scaling1b})} reduces to
\begin{eqnarray}
\label{eq.scaling1b2}
\frac{d\rho_0}{dt}\biggl (S_1(x)+\frac{1}{2}x S'_1(x)\biggr)
&=& \rho_{0}^{2}  \left(S''_1(x) + \frac{d + 1}{x} S'_1(x)
+ x S_1(x) S'_1(x) + d S_1^{2}(x)\right), \nonumber
\end{eqnarray}
which coincides with the equation obtained in the single-species case
\cite{sc}. Setting $\rho_0^{-2}d\rho_0/dt=2$, we find that
\begin{equation}
\label{eq.rho0}
\rho_0(t) = \frac{1}{2}\left(t_{coll} - t\right)^{-1}.
\end{equation}
Thus, the central density diverges in a finite time $t_{coll}$.
Furthermore, the differential equation for the invariant profile can 
be solved analytically \cite{sc} and we get $S_{1}(x)=S_{0}(x)$ where
\begin{equation}
\label{eq.S0}
S_0(x) =\frac{4}{d - 2 + x^2}.
\end{equation}
Using the preceding results, the differential equation determining the
invariant profile of species $2$ is given by the linear second order
differential equation
\begin{equation}
\label{eq.S}
\zeta \mu S''_{2}(x) + \left[\zeta\left(\frac{\mu (d+1)}{x} +
x S_0(x)\right) - x\right]S'_{2}(x) + \left(d \zeta S_0(x) -
\alpha\right)S_{2}(x) = 0.
\end{equation}
For $x \rightarrow + \infty$, we have the asymptotic behavior
\begin{eqnarray}
\label{eq.Slongrange}
S_{2}(x) \sim x^{-\alpha}.
\end{eqnarray}

\begin{figure}
\vskip 0.5cm
\centerline{
\psfig{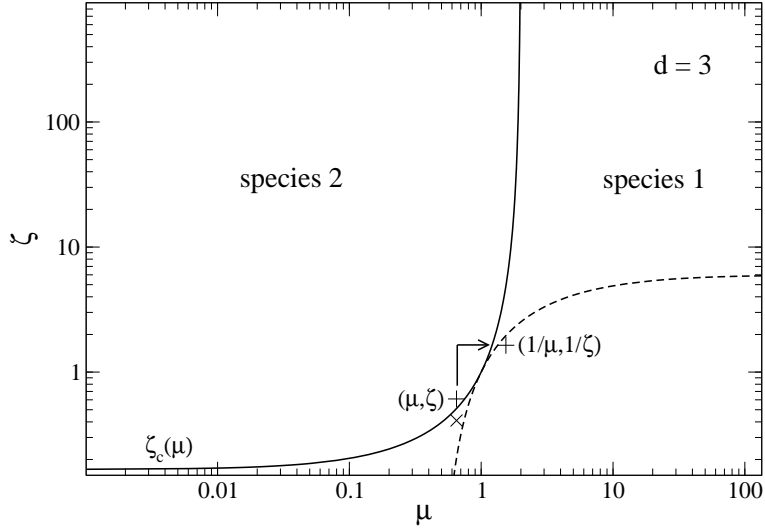}}
\caption{\label{fig.zetac}The critical ratio $\zeta_c$ as a
function of $\mu$ in $d=3$ is plotted in log-log scale. This 
function starts at $\zeta_c(0) = 1/6 = (d-2)/2d$ and diverges 
at $\mu=2$. Below the critical line, species $1$ dominates the 
collapse and above the critical line, species $2$ dominates. 
In that case, our study can still be used with the transformation 
$(\zeta,\mu) \rightarrow (1/\zeta,1/\mu)$. This gives a corresponding 
point located below the dashed curve, corresponding to the function
$1/\zeta_c(1/\mu)$. We show an example illustrating this
transformation when $\zeta>\zeta_c$. The (${\times}$) symbol
represents the point $(0.65,0.41)$ and the ($+$) symbols represent
$(0.65,0.61)$ and $(1/0.65, 1/0.61)$ respectively.}
\end{figure}

Equation~(\ref{eq.S}) can be numerically solved for any couple
$(\mu,\zeta)$. As this equation has been obtained under the assumption
that the exponent $\alpha < 2$, we define a critical ratio of friction
coefficients $\zeta_c(\mu)$ corresponding to the limit of validity of
this hypothesis, {i.e.} $\alpha(\mu,\zeta_c) = 2$ (similarly, we 
define $\mu_c(\zeta)$ such that $\alpha(\mu_c,\zeta) = 2$). 

In the case of a discrete spectrum of particle masses and friction
coefficients, the above calculations can be repeated. One obtains an
equation identical to Eq.~(\ref{eq.S}) for each type of sub-dominant
particles, for which the analysis that we present below has to be
applied.

The critical ratio $\zeta_c(\mu)$ is plotted in
\fig{\ref{fig.zetac}}. The value $\zeta_c(0)$ can be obtained
analytically. Inserting $\mu=0$ and $\alpha=2$ in \eqn{(\ref{eq.S})},
we obtain
\begin{equation}
\label{eq.Smu0zc}
x( \zeta_c S_0(x) - 1)S'_2(x) + (d \zeta_c S_0(x) - 2)S_2(x) = 0.
\end{equation}
This equation must have a solution for any value of $x$. Taking $x=0$, 
we find the necessary condition
\begin{equation}
\label{eq.zetac0}
\zeta_c(0) = \frac{d-2}{2d}.
\end{equation}
Inserting this result in \eqn{(\ref{eq.Smu0zc})}, we find that the 
scaling profile $S_2(x)$ is
\begin{equation}
\label{eq.S2mu0}
S_2(x) = \frac{A}{(d-2)^2/d+x^2},
\end{equation}
where $A$ is an integration constant. The assumption that species $1$
dominates the collapse is valid for $\zeta<\zeta_c(\mu)$. Above this
critical line, the role played by the two species is swapped, and
species $2$ dominates. We can return to the studied situation by
simply changing the index $1 \leftrightarrow 2$. Therefore, if
$(\mu,\zeta)$ belongs to region $2$ in \fig{\ref{fig.zetac}}, this
transformation leads to study the case $(1/\mu,1/\zeta)$ which belongs
to a sub-part of region $1$, below the dashed line.

We have numerically studied this inversion in \fig{\ref{fig.equiv}}. 
We first start with a value of $(\mu,\zeta)$ below the critical line. 
Specifically, we take $\mu=0.65$ (leading to $\zeta_{c}\simeq 0.51$) 
and $\Delta=\zeta-\zeta_c=-0.1$ (point $\times$). In that case, species 
$1$ dominates the collapse : its profile decreases as $\rho_{1}\sim r^{-2}$ 
while the profile of species $2$ decreases as $\rho_{2}\sim r^{-\alpha}$ 
with $\alpha=1.85337<2$ determined by solving numerically the scaling
equation (\ref{eq.S}) with $(\mu,\zeta)$. We then increase the value
of $\zeta$ above the critical line, at the same distance
$\Delta=\zeta-\zeta_c=+0.1$ (point $+$). In that case, we have a
reversal of population. It is now species $2$ that dominates the
collapse : its profile decreases as $\rho_{2}\sim r^{-2}$ while the
profile of species $1$ scales as $\rho_{1}\sim r^{-\alpha'}$. To get
the value of $\alpha'$ from our study, we set $2\rightarrow I$ and
$1\rightarrow II$. We are now in the situation where species I
dominates. Due to this transformation, the new parameters are
$\tilde\mu\equiv m_{I}/m_{II}=1/\mu$ and $\tilde\zeta\equiv
\zeta_{I}/\zeta_{II}=1/\zeta$. Then, $S_{I}=S_{2}$ is given by 
\eqn{(\ref{eq.S0})} and $S_{II}=S_{1}$ is solution of \eqn{(\ref{eq.S})} 
with $\tilde\mu$ and $\tilde\zeta$. The numerical solution of this
scaling equation gives $\alpha'=1.74238$. We note that $\alpha'\neq
\alpha$ so that the slope of the function $\alpha(\Delta)-2$ is
discontinuous as $\Delta\rightarrow 0$.

\begin{figure}
\vskip0.5cm
\centerline{
\psfig{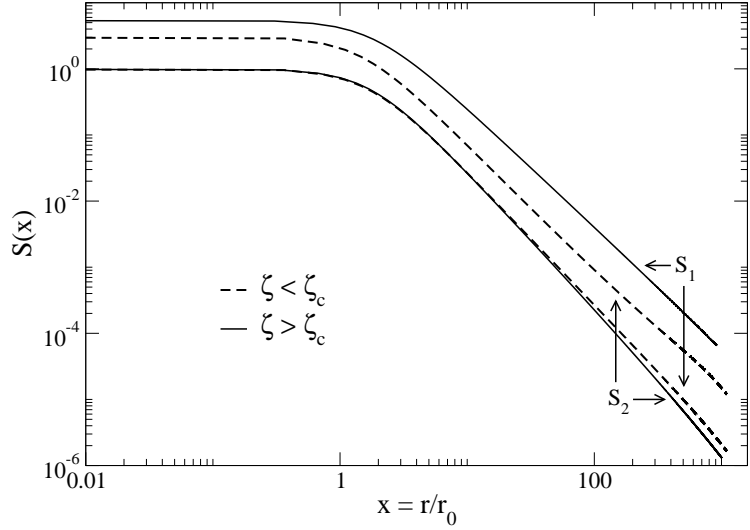}}
\caption{\label{fig.equiv}We plot the scaling profiles $S_1$ and $S_2$
for $\mu=0.65$ and for different values of $\zeta$. We take $\zeta
=0.41<\zeta_{c}=0.511070$ (dashed lines) and $\zeta =0.61>\zeta_{c}$
(solid lines). This corresponds to the points marked ${\times}$ and
$+$ in \fig{\ref{fig.zetac}}.  For $\zeta>\zeta_c$, the exponent
$\alpha'$ is obtained from
\eqn{(\ref{eq.S})} using the equivalent point $(1/\mu,1/\zeta)$.}
\end{figure}

For $\mu\rightarrow 2$, the critical ratio $\zeta_{c}(\mu)$
diverges. Therefore, for $\mu > 2$, species $1$ always dominates 
the collapse whatever the value of $\zeta$. It is possible to
show the signature of this phenomenon analytically. Assuming 
$\alpha=2$ and $\zeta_c \rightarrow +\infty$, \eqn{(\ref{eq.S})} 
reduces to
\begin{equation}
\label{eq.zcdv}
S''_2 + \left(\frac{d+1}{x}+\frac{x S_0}{\mu}\right)S'_2 + \frac{d
S_0}{\mu} S_2 = 0.
\end{equation}
For large $x$, the profile $S_2(x)$ should decay as $x^{-2}$, which
immediately implies that $\mu=2$. Equation (\ref{eq.zcdv}) for
$S_2(x)$ can then be solved in terms of hypergeometric functions.  On
the other hand, considering a perturbation expansion $d \rightarrow
+\infty$ (see \sec{\ref{ssec.pertd}}), we can obtain the analytical
expression
\begin{eqnarray}
\label{eq.zetainf}
\zeta_c(\mu) = \frac{1}{2-\mu}, \qquad  (d \rightarrow +\infty).
\end{eqnarray}
We note that, in this limit, $\zeta_c(0)=1/2$ in agreement with the
exact result (\ref{eq.zetac0}). We can also obtain an approximate
expression of the profile $S_{2}(x)$ for $d\rightarrow +\infty$ (see
\sec{\ref{ssec.pertd}}).

\end{subsection}

\begin{subsection}{Perturbation expansion for $d \rightarrow +\infty$}
\label{ssec.pertd}

We shall first obtain the expression of the scaling exponent $\alpha$
for $d\rightarrow +\infty$. We shall see that the resulting expression
applied for $d=3$ already provides a good approximation of the exact solution.
We use a method similar to that developed in \cite{sc} in a slightly different
context. Equation (\ref{eq.S}) can be formally written as a first order
differential equation (writing $S''_{2} = (S''_{2}/S'_{2}) S'_{2}$)
depending on $x$, $S_0$ and $S''_{2} /S'_{2}$,
\begin{equation}
\label{eq.S'S}
\frac{S'_{2}}{S_{2}} = \frac{d\zeta S_0(x) - \alpha}
{x - \zeta\left[x S_0(x) + \mu\left(\frac{d + 1}{x} + \frac{
S''_{2}}{S'_{2}}\right)\right]}.
\end{equation}
The term $d \zeta S_0(x) - \alpha$ vanishes for a particular $x$, noted
$x_0 \ne 0$, whereas the ratio $S'_{2}/S_{2}$ cannot vanish. This implies
that the denominator in \eqn{(\ref{eq.S'S})} must be equal to zero for 
$x=x_0$. Using \eqn{(\ref{eq.S0})}, $x_{0}$ is explicitly given by
\begin{equation}
\label{eq.x0}
x_0 = \sqrt{d \left(\frac{4 \zeta}{\alpha} - 1\right) + 2}.
\end{equation}
The condition that the denominator vanishes for this value can be written
\begin{equation}
\label{eq.S'S1}
x_0^2 - \zeta \left[\mu\left(d + 1 + x_0 \frac{S''_{2}(x_0)}{S'_{2}(x_0)}
\right) + x_0^2 S_0(x_0)\right] = 0.
\end{equation}
In the sequel, it is more convenient to work with the variable 
$u = x^2$. In terms of this variable, \eqns{(\ref{eq.x0})} and
(\ref{eq.S'S1}) become
\begin{eqnarray}
\label{eq.u0}
&\displaystyle u_0= d \left(\frac{4\zeta}{\alpha} - 1\right) + 2,& \\
\label{eq.S'S2}
&\displaystyle u_0 - \zeta \left[\mu\left(d + 2 + 2 u_0\frac{S''_{2}(u_0)}
{S'_{2}(u_0)}\right) + u_0 S_0(u_0)\right]
= 0.&
\end{eqnarray}
Using \eqns{(\ref{eq.S0})} and (\ref{eq.u0}), \eqn{(\ref{eq.S'S2})} can be
rewritten
\begin{eqnarray}
\label{eq.S'S3}
d \left(\frac{4 \zeta}{\alpha} - 1 - \zeta \mu \right) + 2\left[1 - \zeta
\left(1 + \mu + \mu d \left(\frac{4 \zeta}{\alpha} - 1\right)\frac{S''_{2}
(u_0)}{S'_{2}(u_0)} + \frac{\alpha}{2}\right)\right] \nonumber \\
 - 2 \left(\zeta \mu \frac{S''_{2}(u_0)}{S'_{2}(u_0)} + \frac{\alpha}{d}
\right) = 0.
\end{eqnarray}
In the limit $d \rightarrow + \infty$,  keeping only the dominant terms in
the above equation, we obtain  $\zeta \mu + 1 - 4\zeta/\alpha = 0$ from 
which we derive the zeroth order expression of $\alpha$
\begin{equation}
\alpha = \frac{4 \zeta}{1 + \zeta \mu}.
\label{zer}
\end{equation}
From this last equation, taking $\alpha=2$, we get \eqn{(\ref{eq.zetainf})}. 
Substituting this result in \eqn{(\ref{eq.S'S})} and keeping only the leading 
terms for $d\rightarrow +\infty$, we get
\begin{equation}
\label{eq.S'So1}
\frac{S'_{2}(u)}{S_{2}(u)} = - \frac{2\zeta}{(d+u)(1+\zeta\mu)}.
\end{equation}
This equation is easily integrated and  leads to the first approximation of
$S_{2}(x)$ in the large $d$ limit
\begin{equation}
\label{eq.So1}
S_{2}(x) = \frac{A}{(d + x^2)^{\frac{2\zeta}{1+\zeta\mu}}}
\end{equation}
where $A$ is an integration constant which cannot be determined
explicitly at this order. We are now able to obtain the next order
correction of $\alpha$. Let us write
\begin{equation}
\alpha = \frac{4\zeta}{1+\zeta\mu} + \frac{\alpha_1}{d}.
\end{equation}
Inserting this expression in \eqn{(\ref{eq.S'S3})}, considering the limit
$d\rightarrow +\infty$ and using \eqns{(\ref{eq.S0})} and (\ref{eq.So1}),
we finally obtain
\begin{equation}
\alpha_1 = -\frac{8\zeta\left(2\zeta^2\mu-\zeta\mu-1\right)}{\left(1+\zeta
\mu\right)^4}.
\end{equation}
This leads to the approximate expression of $\alpha$ to order $1/d$,
\begin{equation}
\label{eq.alphalarged}
\alpha = \frac{4\zeta}{1 + \zeta\mu}\left[1-\frac{2(2\zeta^2\mu-\zeta\mu-1)}
{d(1+\zeta\mu)^3} + O\left(\frac{1}{d^2}\right)\right].
\end{equation}
This expression is valid for arbitrary values of $\mu$ and $\zeta$
such that $\alpha < 2$.

\end{subsection}

\begin{subsection}{Perturbation expansion for $\zeta \sim 1$
and $\mu \sim 1$}
\label{ssec.pert1}

We now consider the case of weak asymmetry $\mu \sim 1$ and $\zeta
\sim 1$ between the two species for any dimension $d$. In that case,
$S_{2}(x)$ will be close to $S_0(x)$ and $\alpha$ will be close to $2$.
We set
\begin{equation}
\label{eq.smallpert}
\zeta = 1 - \epsilon \; , \; \mu = 1 + \eta \; , \;  \alpha = 2 - \epsilon
\alpha_{\zeta} - \eta \alpha_{\mu},
\end{equation}
\begin{equation}
\label{eq.Ssmallpert}
S_{2}(x) = S_0(x) \left[1 + \epsilon g_{\zeta}(x) + \eta g_{\mu}(x)\right] ,
\end{equation}
with $\epsilon,\eta \ll 1$. Substituting this expansion in \eqn{(\ref{eq.S})}, 
it is found that the functions $g_{\zeta}(x)$ and $g_{\mu}(x)$ satisfy the 
{\it first order} differential equations (for their derivatives)
\begin{eqnarray}
\label{eq.g''zeta}
g''_{\zeta}(x) + \left(\frac{d+1}{x}-x\right) g'_{\zeta}(x) = \frac{2(d-2)}
{d-2+x^2} - \alpha_{\zeta}, \\
\label{eq.g''mu}
g''_{\mu}(x) + \left(\frac{d+1}{x}-x\right) g'_{\mu}(x) = \frac{2(d-2)(x^2+
d+2)}{(d-2+x^2)^2} - \alpha_{\mu}.
\end{eqnarray}
We shall discuss these equations separately.

\begin{subsubsection}{The case $\mu = 1$}
\label{sssec.mu1}

\begin{figure}
\vskip0.5cm
\centerline{
\psfig{figure=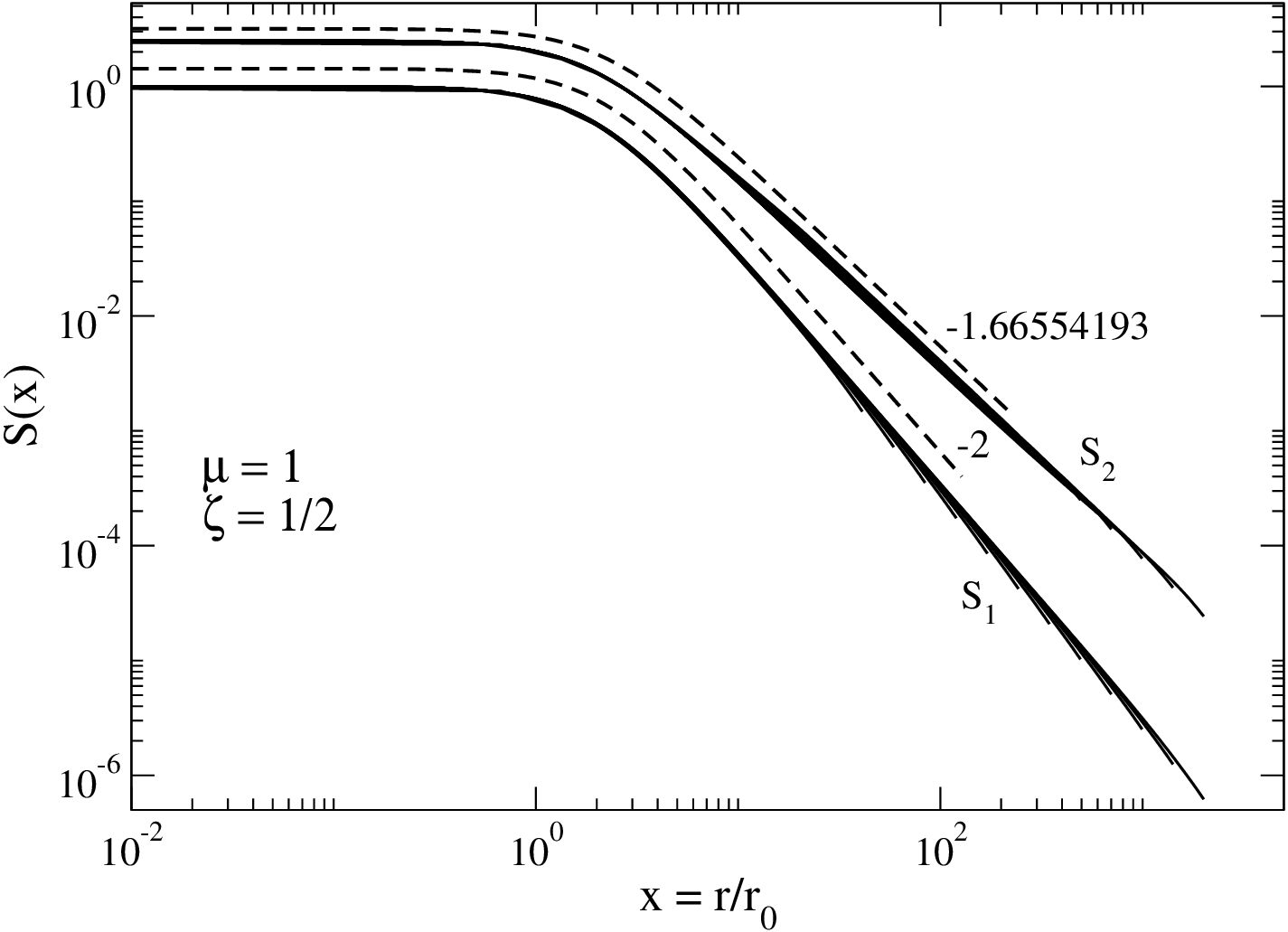,angle=0,height=7cm}}
\caption{\label{fig.scazeta} The resolution of the time-dependent 
equations (\ref{eq.intmass}) shows that the evolution is self-similar. 
We fix for the simulation: $\mu = 1$, $d = 3$, $T = 0.2$, $\zeta = 0.5$ 
and $M_1= M_2 = 0.5$. For $t\rightarrow t_{coll}$, the rescaled densities 
converge to the invariant profiles $S_1(x)$ and $S_2(x)$ predicted by the 
theory. The profile of species $1$ is the same as in the one-component 
problem : $S_1(x) = S_0(x) \sim x^{-2}$. The profile of species $2$ has been 
obtained by solving \eqn{(\ref{eq.S})} numerically : $S_2(x) \sim x^{-\alpha}$ 
with $\alpha = 1.66554193$.}
\end{figure}

We first consider the case $\mu=1$. Equations (\ref{eq.intmass}) have
been solved numerically for $\zeta=1/2$ and the corresponding scaling
profiles are plotted in \fig{\ref{fig.scazeta}. The numerical results
lead to the predicted exponents : at large $x$, $S_1(x) \sim x^{-2}$
and $S_2(x) \sim x^{-\alpha}$, where $\alpha$ is calculated using
\eqn{(\ref{eq.S})}}. We now consider the weak asymmetry limit 
$\zeta=1-\epsilon$ with $\epsilon\ll 1$ for $\mu=1$ (the condition 
$\alpha<2$ imposes $\epsilon>0$). Then, $S_{2}(x)=S_{0}(x)[1+\epsilon 
g_{\zeta}(x)]$ where $g_{\zeta}(x)$ is the solution of 
\eqn{(\ref{eq.g''zeta})}. This equation can be integrated once leading to
\begin{equation}
\label{eq.g'zeta}
g'_{\zeta}(x) = x^{-(d+1)} e^{x^2/2}
\int_0^x y^{d+1} e^{-y^2/2} \left(\frac{2(d-2)}{d-2+y^2} -
\alpha_{\zeta}\right)\,dy.
\end{equation}
The integration constant has been determined so as to satisfy the
boundary condition $g'_{\zeta}(0)=0$. Now, the condition that
$g'_{\zeta}(x) \rightarrow 0$ as $x\rightarrow +\infty$, leads to
an exact expression of $\alpha_{\zeta}$. As $x^{-(d+1)}e^{x^2/2}
\rightarrow + \infty$ for $x \rightarrow + \infty$, the integral in
\eqn{(\ref{eq.g'zeta})} has to vanish at large $x$. This yields
\begin{equation}
\label{eq.alphazeta}
\alpha_{\zeta}(d) = \frac{\int_0^{+\infty} y^{d+1} e^{-y^2/2}
\frac{2(d-2)}{d-2+y^2}\,dy}{\int_0^{+\infty} y^{d+1} e^{-y^2/
2}\,dy} \ge 0.
\end{equation}
Note that the integrals can be expressed in terms of $\Gamma$ functions.
Rewriting \eqn{(\ref{eq.g'zeta})} in the form
\begin{equation}
g'_{\zeta}(x) = - x^{-(d+1)} e^{x^2/2} \int_x^{+\infty} y^{d+1} e^{-y^2/2}
\left(\frac{2(d-2)}{d-2+y^2} - \alpha_{\zeta}\right)\,dy,
\end{equation}
we derive the large $x$ behaviors
\begin{eqnarray}
g'_{\zeta}(x) \sim \frac{\alpha_{\zeta}}{x} &,& g_{\zeta}(x) \sim
\alpha_{\zeta} \ln x.
\end{eqnarray}
We can also carry an expansion of $\alpha_{\zeta}(d)$ in powers of
$d^{-1}$ in the limit $d\rightarrow +\infty$. Using the saddle point
method in \eqn{(\ref{eq.alphazeta})} around the point $y = \sqrt{d + 1}$, 
we obtain
\begin{eqnarray}
\label{eq.alphazetalarged}
\alpha_{\zeta}(d) = 1 - \frac{3}{2d} - \frac{1}{4d^2} - \frac{15197}
{25920d^3} - \frac{266999}{311040d^4} + O\left(\frac{1}{d^5}\right).
\end{eqnarray}
We can check that the first terms of this expansion reproduce those
given by \eqn{(\ref{eq.alphalarged})} for $\mu = 1$ and $\zeta=1-
\epsilon$.

\end{subsubsection}

\begin{subsubsection}{The case $\zeta = 1$}
\label{sssec.zeta1}

\begin{figure}
\vskip0.5cm
\centerline{
\psfig{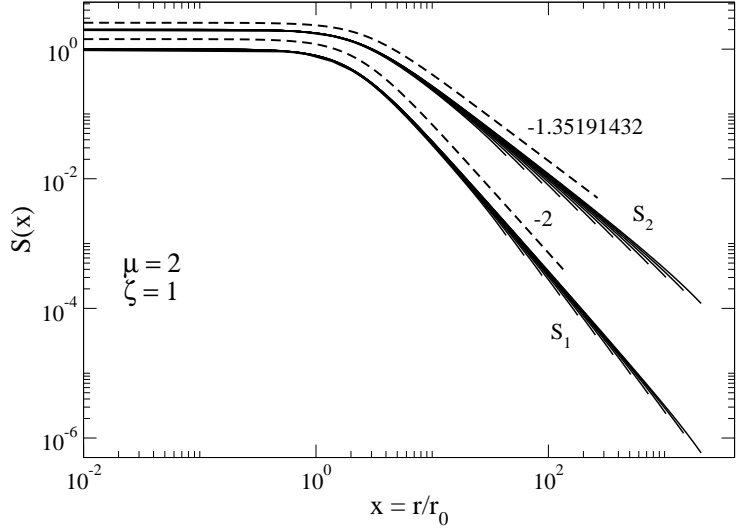}}
\caption{\label{fig.scamu} The resolution of the time-dependent 
equations (\ref{eq.intmass}) shows that the evolution is
self-similar. We fix for the simulation : $\zeta=1$, $d = 3$, $T =
0.2$, $\mu = 2$ and $M_1 = M_2 = 0.5$. For $t\rightarrow t_{coll}$,
the rescaled densities converge to the invariant profiles $S_1(x)$ and
$S_2(x)$ predicted by the theory.  The profile of species $1$ is the
same as in the one-component problem : $S_1(x) = S_0(x) \sim
x^{-2}$. The profile of species $2$ has been obtained by solving
\eqn{(\ref{eq.S})} numerically : $S_2(x) \sim x^{-\alpha}$ with
$\alpha = 1.35191432$. Note that the large $d$ expansion
Eq.~(\ref{zer}) leads to $\alpha=4/3$ in fair agreement with the exact
numerical result.}
\end{figure}

We now consider the case $\zeta=1$. Equations (\ref{eq.intmass}) have
been solved numerically for $\mu=2$ and the corresponding scaling
profiles are plotted in \fig{\ref{fig.scamu}}. They converge to
the invariant profiles predicted by theory. We now consider the weak
asymmetry limit $\mu=1+\eta$ with $\eta\ll 1$ for $\zeta=1$ (the
condition $\alpha<2$ imposes $\eta>0$). Then, $S_{2}(x)=S_{0}(x)[1+
\eta g_{\mu}(x)]$ where $g_{\mu}(x)$ is solution of \eqn{(\ref{eq.g''mu})}. 
Following a procedure similar to that exposed previously, we get the 
following expression of $\alpha_{\mu}$,
\begin{equation}
\label{eq.alphamu}
\alpha_{\mu}(d) = \frac{\int_0^{+\infty} y^{d+1} e^{-y^2/2}
\frac{2(d-2)(y^2+d+2)}{(d-2+y^2)^2}\,dy}{\int_0^{
+\infty}y^{d+1} e^{-y^2/2}\,dy} \ge 0.
\end{equation}
and the asymptotic behaviors
\begin{eqnarray}
g'_{\mu}(x) \sim \frac{\alpha_{\mu}}{x} &,& g_{\mu}(x) \sim \alpha_{
\mu}\ln x.
\end{eqnarray}
The large $d$ expansion of \eqn{(\ref{eq.alphamu})} is
\begin{eqnarray}
\label{eq.alphamularged}
\alpha_{\mu}(d) = 1 + \frac{1}{2d} - \frac{5}{4d^2} - \frac{9}{8
d^3} -\frac{23}{16d^4} + O\left(\frac{1}{d^5}\right) &,& d
\rightarrow + \infty,
\end{eqnarray}
and the first terms of this expansion reproduce those of
\eqn{(\ref{eq.alphalarged})} for $\zeta=1$ and $\mu= 1+\eta$.
The exact values of $\alpha_{\zeta}(d)$ and $\alpha_{\mu}(d)$
along with their $O(d^{4})$ expansions are plotted in 
\fig{\ref{fig.alphazm}}. For $d = 3$, the exact values 
are $\alpha_{\zeta}(3) = 0.437119695$ and $\alpha_{\mu}(3) =
0.940162135$.

\begin{figure}
\vskip0.5cm
\centerline{
\psfig{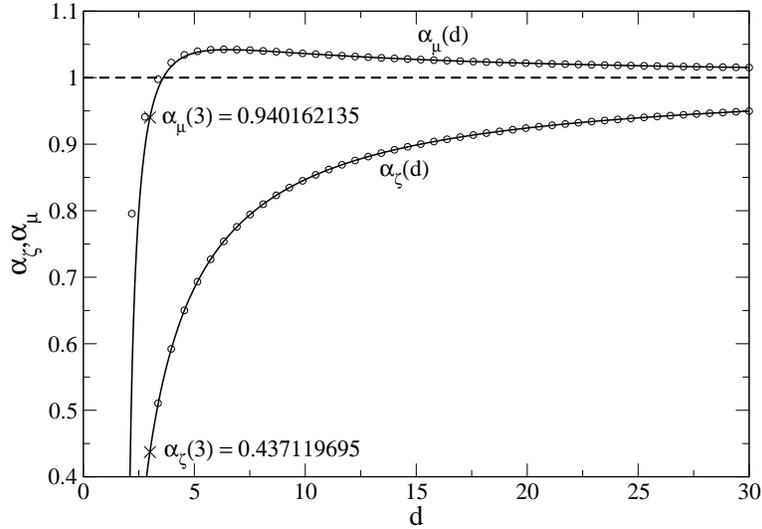}}
\caption{\label{fig.alphazm} Numerical calculation of $\alpha_{\zeta}$ 
and $\alpha_{\mu}$ given by \eqns{(\ref{eq.alphazeta})} and 
(\ref{eq.alphamu}) as a function of the dimension $d$. The
dashed line represents the asymptotic value for $d \rightarrow
+\infty$. The $(\circ)$ symbols represent the large $d$ expansion of
$\alpha_\zeta$ and $\alpha_\mu$ to order $d^{-4}$ given in
\eqns{(\ref{eq.alphazetalarged})} and (\ref{eq.alphamularged}).}
\end{figure}

\end{subsubsection}

\end{subsection}

\begin{subsection}{Other perturbation expansions}
\label{ssec.otherpert}

We now consider perturbation expansions of \eqn{(\ref{eq.S})} for small 
and large values of $\mu$ and $\zeta$. For $\mu \rightarrow 0$ and 
$\zeta<\zeta_c$, \eqn{(\ref{eq.S})} reduces to
\begin{equation}
\label{eq.Smu0}
x( \zeta S_0(x) - 1)S'_2(x) + (d \zeta S_0(x) - \alpha)S_2(x) = 0.
\end{equation}
Considering the value $x=0$, we get $d \zeta S_0(0)-\alpha=0$ leading to
\begin{equation}
\label{eq.alphamu0}
\alpha= \frac{4d\zeta}{d-2}.
\end{equation}
Then, the scaling profile is given by 
\begin{equation}
S_2(x)= \frac{A}{(d-2-4 \zeta +x^2)^{2d\zeta/(d-2)}}.
\end{equation}
We now wish to examine the limit $\mu \rightarrow + \infty$. We assume 
that $\alpha \sim 1/\mu$ and check this scaling a posteriori. Using the 
fact that $S_{2}\sim x^{-\alpha}$ for $x\rightarrow +\infty$ and comparing 
terms of order $x^{-\alpha-2}$ in \eqn{(\ref{eq.S})}, we find that 
\begin{equation}
\label{eq.alphamuinf}
\alpha = \frac{4}{\mu}.
\end{equation}
We note that this expression is independent on $\zeta$. The scaling
profile obtained from \eqn{(\ref{eq.S})} can be expressed in terms of
hypergeometric functions.

In the limit $\zeta\rightarrow 0$, \eqn{(\ref{eq.S})} simplifies into 
\begin{equation}
\label{eq.Szeta0}
- x S'_2(x) + (d \zeta S_0(x) - \alpha) S_2(x) = 0.
\end{equation}
Considering the value $x=0$, we get
\begin{equation}
\label{eq.alphazeta0}
\alpha= \frac{4d\zeta}{d-2}.
\end{equation}
Then, the scaling profile $S_2(x)$ takes the form
\begin{equation}
\label{eq.S2zeta0}
S_2(x) = \frac{A}{(d-2+x^2)^{2d\zeta/(d-2)}}.
\end{equation}
We now wish to examine the limit $\zeta \rightarrow + \infty$ and  $\mu>2$. 
Using the fact that $S_{2}\sim x^{-\alpha}$ for $x\rightarrow +\infty$ and 
comparing terms of order $x^{-\alpha-2}$, we find that 
$\mu\alpha^{2}-(4+d\mu)\alpha+4d=0$ leading to $\alpha=d$ or $\alpha=4/\mu$. 
Since $\alpha<2$, we get
\begin{equation}
\label{eq.alphazetainf}
\alpha = \frac{4}{\mu}.
\end{equation}
Figure \ref{fig.az1m1} shows the functions $\alpha(\zeta=1,\mu)$ and
$\alpha(\zeta,\mu=1)$ obtained by solving \eqn{(\ref{eq.S})} in $d=3$
and compares these numerical results with the asymptotic expansions
obtained previously. For $\zeta$ and $\mu$ close to 1, the slope of
the function $\alpha(\zeta,\mu)$ is given by \eqns{(\ref{eq.alphamu})}
and (\ref{eq.alphazeta}). This figure also confirms the asymptotic
expressions (\ref{eq.alphamu0}) and (\ref{eq.alphazeta0}) obtained for
$\mu \rightarrow + \infty$ and $\zeta \rightarrow 0$.

\begin{figure}
\vskip0.5cm
\centerline{
\psfig{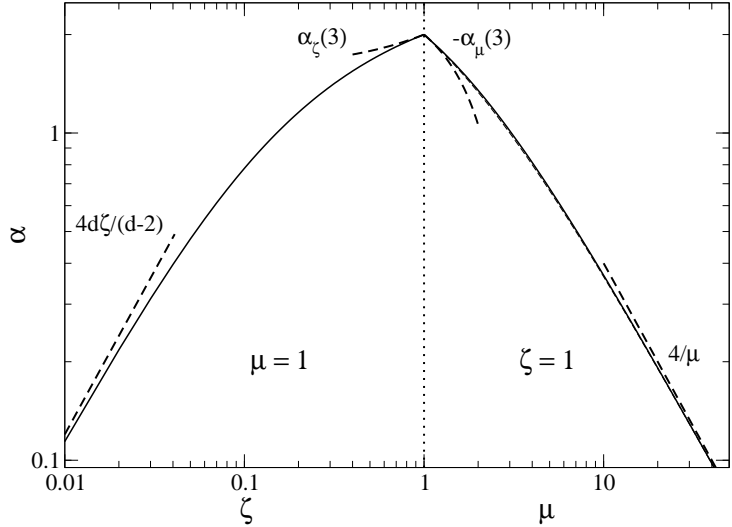}}
\caption{\label{fig.az1m1}The exponent $\alpha$ is plotted for 
$\mu=1,\zeta \le 1$ and for $\mu \ge 1,\zeta=1$ in $d=3$. The dashed
lines give the different asymptotic behaviors obtained
analytically. Finally, the dotted line for $\zeta=1$ corresponds to
the result Eq.~(\ref{zer}) of the large $d$ expansion, and is in
excellent agreement with the exact value of $\alpha$ (the two curves
are almost indistinguishable).}
\end{figure}

\end{subsection}

\end{section}

\begin{section}{Conclusion}

In this paper, we have extended previous studies on the thermodynamics
of self-gravitating particles in $d$-dimensions to the case of
multi-components systems. Our static study applies both to the
microcanonical (fixed $E$) and canonical (fixed $T$) ensembles. Thus,
it describes ordinary stellar systems (like globular clusters)
\cite{bt}, self-gravitating Brownian particles \cite{prs} and
bacterial populations \cite{murray,crrs}. We have investigated how the
critical energy (Antonov point) and the critical temperature (Jeans
point) depend on the parameters. If we take as a reference a
single-species system with particles of mass $m_{2}$ and add particles
of mass $m_{1}$ (while removing some particles of mass $m_{2}$ so as
to keep the total mass $M$ fixed), we find that the critical
temperature is increased if $m_{1}>m_{2}$ and decreased if
$m_{1}<m_{2}$ (an analytical estimate of the critical temperature has
been obtained in $d>2$ using the Jeans swindle). By contrast, the
critical energy is always increased with respect to the single species
case in $d>3$. For given ratio $\mu=m_{1}/m_{2}$, it presents a
maximum at a certain value of $\chi=M_{1}/M_{2}\simeq 0.7$
(\fig{\ref{fig.etalambda13d}}.b). This maximum energy increases
roughly linearly with $\mu$ (\fig{\ref{fig.Lcmin}}). As in the
one-component case, two-dimensional systems require a specific
attention. In $d=2$, there is no collapse in the microcanonical
ensemble but there is a collapse in the canonical ensemble below a
critical temperature. We have obtained this critical temperature
analytically. For $\mu\le 2$ and for ($\mu>2$ and
$\chi<\chi_{*}=1/(\mu-2)$), the two species of particles form a Dirac
peak at $T=T_c$ and the expression (\ref{eq.etac1}) of the critical
temperature can be obtained from the Virial theorem. For ($\mu>2$ and
$\chi>\chi_{*}=1/(\mu-2)$), only the heaviest particles form a Dirac
at $T=T_{c}$ and the expression of the critical temperature
(\ref{eq.etac2}) is different.

We have also studied the dynamics of self-gravitating Brownian
particles (and bacterial populations) in the framework of the
two-species Smoluchowski-Poisson system. This corresponds to the
canonical ensemble. For $T<T_c$, there is no equilibrium state and the
system collapses. Looking for self-similar solutions, we have shown
that one species dominates the other and collapses as in the
single-species problem with a scaling profile $\rho(r)\sim
r^{-2}$. The selection of the dominant species is non-trivial. For
$\mu>2$, the dominant species is always the one with the heaviest
particles.  For $\mu<2$, the selection depends on the ratio $\zeta$ of
friction parameters as shown in \fig{\ref{fig.zetac}} (for $\zeta=1$,
the species with heaviest particles always dominates the
collapse). The scaling profile of the ``slaved'' species decays with
an exponent $\alpha<2$ depending on $d$, $\mu$ and $\chi$. This
exponent can be calculated numerically by solving
\eqn{(\ref{eq.S})}. We have also given several asymptotic expansions,
see \eqns{(\ref{eq.alphalarged})}, (\ref{eq.alphazeta}),
(\ref{eq.alphamu}), (\ref{eq.alphamu0}), (\ref{eq.alphamuinf}),
(\ref{eq.alphazeta0}) and (\ref{eq.alphazetainf}).

The generalization of our approach to a {\it continuous} spectrum of masses
and friction coefficients does not look straightforward. Let us focus
on the simpler case of identical friction coefficients. The precise
form of the mass spectrum is certainly highly relevant. In particular,
we expect that the value of the minimum and maximum masses (possibly 0
and $+\infty$) is crucial. In addition, the behavior of the
distribution near the largest mass (for instance $p(m)\sim (m-m_{\rm
max})^{-\gamma}$ if $m_{\rm max}$ is finite ($\gamma<1$), $p(m)\sim
m^{-\gamma}$ otherwise ($\gamma>1$)) is certainly an important
ingredient.  However, in the case of a bounded distribution of mass
without extravagant singularities, we expect that the results obtained
in the present paper will qualitatively hold: the heaviest particles
will scale as in the one species case, while lighter particles will
scale with a mass dependent exponent less than 2.

\end{section}


\appendix

\begin{section}{Derivation of the mean-field equations}
\label{app.meanfield}

In this Appendix, we show that the mean-field approximation used in
our study is exact in a proper thermodynamic limit (see \cite{chav} in the single species case). We consider the
case of self-gravitating Brownian particles described by the stochastic
equations (\ref{eq.stocha}). The proper statistical ensemble for this
system is the canonical ensemble. At equilibrium, 
the $N$-body distribution function is given by
\begin{equation}
\label{eq.PNrv}
P_N(\textbf{r}_1,\textbf{v}_1,{...},\textbf{r}_N,\textbf{v}_N) = \frac{1}
{Z_T}e^{-\beta H(\textbf{r}_1,\textbf{v}_1,{...},\textbf{r}_N,\textbf{v}_N)},
\end{equation}
where $Z_T$ is the partition function (normalization constant) and $H$ is the
Hamiltonian
\begin{eqnarray}
\label{eq.hamiltonian}
H &=& \frac{1}{2}\sum_{i = 1}^{N} m_i v_i^2 + \sum_{i<j} m_{i}m_{j}
u(\textbf{r}_i - \textbf{r}_j)=K+U,
\end{eqnarray}
where $u(\textbf{r}_i - \textbf{r}_j) = u_{ij} =
- {G}/\lbrack (d-2)|\textbf{r}_i - \textbf{r}_j |^{(d-2)}\rbrack$ is the
gravitational potential. From \eqn{(\ref{eq.hamiltonian})}, it is clear
that the velocity distribution is Gaussian. We shall therefore restrict
ourselves to the configurational part
\begin{equation}
\label{eq.PNr}
P_N(\textbf{r}_1,{...},\textbf{r}_N) = \frac{1}{Z} e^{- \beta
U(\textbf{r}_1,{...},\textbf{r}_N)}.
\end{equation}
We introduce the density probability for particle $i$ of species $\alpha$
to be at $\textbf{r}_i$, namely
\begin{equation}
\label{eq.P1alpha}
P_1^{(\alpha)}(\textbf{r}_i) = \int P_N(\textbf{r}_1,{...},
\textbf{r}_N)\,\prod_{j \ne i} d{\bf r}_j.
\end{equation}
Similarly, we define the density probability to find particle $i$ of
species $\alpha$ at $\textbf{r}_i$ and  particle $j$ of species
$\alpha'$ at $\textbf{r}_j$,
\begin{equation}
\label{eq.P2aa'}
P_2^{(\alpha \alpha')}(\textbf{r}_i,\textbf{r}_j) = \int
P_N(\textbf{r}_1,{...},\textbf{r}_N)\,\prod_{k \ne i,j} d{\bf r}_k.
\end{equation}
The total density of particles at $\textbf{r}$ is given by
$\rho(\textbf{r}) = \sum_i m_i \delta\left(\textbf{r}-\textbf{r}_i\right)$.
Its mean value $\left<\rho (\textbf{r})\right>
= \sum_i \int m_i \delta\left(\textbf{r}-\textbf{r}_i\right)
P_N(\textbf{r}_1,{...},\textbf{r}_N)\,\prod_j d{\bf r}_j$ can be
written
\begin{equation}
\label{eq.meanrho}
\left<\rho(\textbf{r})\right>
= \sum_{\alpha = 1}^X \sum_{i \in I_{\alpha}}
\int m_{\alpha} \delta\left(\textbf{r}-\textbf{r}_i\right)
P_1^{(\alpha)}(\textbf{r}_i)\,d{\bf r}_i
= \sum_{\alpha = 1}^{X} N_{\alpha} m_{\alpha} P_1^{(\alpha)}(\textbf{r})
= \sum_{\alpha = 1}^{X} {\left<\rho_{\alpha}(\textbf{r})\right>},
\end{equation}
where we have defined $I_{\alpha} = \left[g_{\alpha - 1}N +
1,g_{\alpha}N\right]$ with $g_0 = 0$ and $g_X = 1$ as the interval of
indices labeling particles of species $\alpha$. In the following, we
shall work with the mean density.  Therefore, we drop the brackets
$\left<\cdot\right>$ in order to simplify the notations.
Taking the derivative of \eqn{(\ref{eq.PNrv})} with respect to 
${\bf r}_{i}$, we get
\begin{equation}
\label{eq.dPNdr1}
\frac{\partial P_N}{\partial \textbf{r}_i} = - \beta P_N \frac{\partial U}
{\partial \textbf{r}_i} = - \beta \sum_{j\neq i} P_N \ m_{i}m_{j}
\frac{\partial u_{ij}}{\partial \textbf{r}_i}.
\end{equation}
Assuming that $i\in I_{\alpha}$ and integrating over the other variables, 
we find that
\begin{eqnarray}
\label{eq.dP1dri}
\frac{\partial P_1^{(\alpha)}}{\partial \textbf{r}_i} = - \beta \sum_{j\neq i} \int P_N
\, m_{i}m_{j}\frac{\partial u_{ij}}{\partial \textbf{r}_i}\,\prod_{k\neq i} d{\bf r}_k
&=& - \beta \left(N_{\alpha} - 1\right) \int P_2^{(\alpha\alpha)}(\textbf{r}_i,\textbf{r}_2)
m_{\alpha}^{2}\frac{\partial u_{i2}}{\partial \textbf{r}_i}\,d{\bf r}_2 \nonumber \\
&&- \beta \sum_{\alpha'\neq \alpha} N_{\alpha'} \int P_2^{(\alpha \alpha' )}
(\textbf{r}_i,\textbf{r}_2)m_{\alpha}m_{\alpha'}\frac{\partial u_{i2}}
{\partial \textbf{r}_i}\,d{\bf r}_2.\qquad\qquad
\end{eqnarray}
Similarly, we can write an equation 
for $P_2^{(\alpha \alpha' )}(\textbf{r}_i,\textbf{r}_j)$
by integrating \eqn{(\ref{eq.PNrv})} over $N-2$ variables. Then 
writing
\begin{equation}
P_2^{(\alpha \alpha')}(\textbf{r}_i,\textbf{r}_j) =  P_1^{(\alpha)}(\textbf{r}_i)
P_1^{(\alpha')}(\textbf{r}_j)+P_2^{'(\alpha \alpha')}(\textbf{r}_i,\textbf{r}_j),
\end{equation}
we can show (see \cite{chav} in the single species case) that the cumulating function $P'_{2}$ is of order
$N_{\alpha}^{-1}$ in the limit 
\begin{equation}
\label{tl}
N_{\alpha}\rightarrow +\infty \quad {\rm with \ fixed}\quad
\eta_{\alpha}=\frac{\beta GM_{\alpha}m_{\alpha}}{R^{d-2}} \quad {\rm  and}
\quad\mu_{\alpha}=\frac{m_{\alpha}}{m_{X}}.
\end{equation}
Thus, in this proper thermodynamic limit, we can make the mean-field approximation
\begin{equation}
\label{eq.meanfield}
P_2^{(\alpha \alpha')}(\textbf{r}_i,\textbf{r}_j)
= P_1^{(\alpha)}(\textbf{r}_i) P_1^{(\alpha')}(\textbf{r}_j),
\end{equation}
which consists in neglecting the correlations and replacing the
two-body distribution function by a product of two one-body
distribution functions. Inserting this expression in
\eqn{(\ref{eq.dP1dri})}, we get
\begin{equation}
\frac{\partial P_1^{(\alpha)}}{\partial \textbf{r}_i}
= - \beta m_{\alpha} P_1^{(\alpha)}(\textbf{r}_i) \sum_{\alpha'}
\int N_{\alpha'} P_1^{(\alpha')}(\textbf{r}_2)m_{\alpha'}
\frac{\partial u_{i2}}{\partial \textbf{r}_i}\,d{\bf r}_2.
\end{equation}
Introducing the mean density of each species
\begin{equation}
\rho_{\alpha}({\bf r})=N_{\alpha}m_{\alpha}P_{1}^{(\alpha)}({\bf r}),
\end{equation}
and the gravitational potential
\begin{equation}
\Phi({\bf r})=\sum_{\alpha}\int \rho_{\alpha}({\bf r}')u({\bf r}-{\bf r}')\,
d{\bf r}',
\end{equation}
we can rewrite the above equation in the form
\begin{equation}
\frac{\partial\rho_{\alpha}}{\partial {\bf r}}=-\beta m_{\alpha}\rho_{\alpha}
({\bf r})\nabla\Phi ({\bf r}).
\end{equation}
After integration, we obtain the Boltzmann distribution
\begin{equation}
\label{eq.drhodr}
\rho_{\alpha}({\bf r})=A_{\alpha}e^{-\beta m_{\alpha}\Phi({\bf r})}.
\end{equation}
The mean potential energy $W=\left<U\right>$ is given by
\begin{eqnarray}
W = \frac{1}{2}\sum_{i \ne j} \int m_{i}m_{j}\ u_{ij} P_N\,
\prod_{k = 1}^{N} d{\bf r}_k
&=& \frac{1}{2} \sum_{\alpha} N_{\alpha}(N_{\alpha}-1)m_{\alpha}^2
\int P_2^{(\alpha \alpha)}(\textbf{r}_1,\textbf{r}_2) u_{12}\,
d{\bf r}_1\,d{\bf r}_2 \nonumber\\
&& + \frac{1}{2} \sum_{\alpha \ne \alpha'} N_{\alpha} N_{\alpha'}
m_{\alpha} m_{\alpha'} \int P_2^{(\alpha \alpha')}(\textbf{r}_1,\textbf{r}_2) 
u_{12}\,d{\bf r}_1\,d{\bf r}_2.\qquad\qquad
\end{eqnarray}
Implementing the mean-field approximation (\ref{eq.meanfield}), valid in the
thermodynamic limit, the above expression simplifies into
\begin{eqnarray}
W &=& \frac{1}{2} \sum_{\alpha,\alpha'} N_{\alpha}N_{\alpha'}m_{\alpha}m_{\alpha'}
\int P_{1}^{(\alpha)}({\bf r}_{1}) P_{1}^{(\alpha')}({\bf r}_{2})u_{12}\,d{\bf r}_{1}
\,d{\bf r}_{2},
\end{eqnarray}
which can be finally rewritten as
\begin{equation}
W = \frac{1}{2} \int \rho(\textbf{r}) \Phi(\textbf{r})\,d{\bf r}.
\end{equation}

We now consider the dynamical problem defined by the stochastic equations
(\ref{eq.stocha}). Using the Kramers-Moyal expansion, the Fokker-Planck
equation for the evolution of the $N$-body distribution function
$P_{N}({\bf r}_{1},{\bf v}_{1},...,{\bf r}_{N},{\bf v}_{N},t)$ reads
\begin{equation}
\label{eq.fokker}
\frac{\partial P_N}{\partial t}
+ \sum_{i = 1}^{N} \left(\textbf{v}_i \cdot \frac{\partial P_N}{\partial \textbf{r}_i}
+ {\textbf{F}_i} \cdot \frac{\partial P_N}{\partial \textbf{v}_i}\right)
= \sum_{i = 1}^N \frac{\partial}{\partial \textbf{v}_i} \cdot \left(D_i
\frac{\partial P_N}{\partial \textbf{v}_i}\ + \xi_i P_N \textbf{v}_i\right),
\end{equation}
where $\textbf{F}_i = -(1/ m_{i}) \nabla_{i} U$ is the force by unit of mass (acceleration)
acting on particle $i$. We note that the stationary solution of \eqn{(\ref{eq.fokker})} is
the canonical distribution (\ref{eq.PNrv}) provided that the coefficients of friction and
diffusion are related to each other according to the Einstein formula
\begin{equation}
D_{i}=\frac{\xi_{i}}{\beta m_{i}}.
\end{equation}
Taking $i\in I_{\alpha}$ and integrating over the other variables, we get
\begin{equation}
\frac{\partial P_1^{(\alpha)}}{\partial t}
+ \textbf{v}_{i} \cdot \frac{\partial P_1^{(\alpha)}}{\partial \textbf{r}_i}
+ \int {\textbf{F}_i} \cdot \frac{\partial P_N}{\partial \textbf{v}_i}\,
\prod_{k\neq i} d{\bf r}_k d{\bf v}_k
= \frac{\partial}{\partial \textbf{v}_i} \cdot \left(D_\alpha \frac{\partial P_1^{(\alpha)}}
{\partial \textbf{v}_i} + \xi_\alpha P_1^{(\alpha)} \textbf{v}_i\right).
\end{equation}
Now
\begin{eqnarray}
I \equiv \int {\bf F}_{i} P_{N} \prod_{k\neq i} d{\bf r}_{k}\,d{\bf v}_k
&=& - \int (N_{\alpha}-1)m_{\alpha}\frac{\partial u_{i2}}{\partial {\bf r}_{i}}
P_{2}^{(\alpha\alpha)}({\bf r}_{i},{\bf v}_{i},{\bf r}_{2},{\bf v}_{2},t)\,
d{\bf r}_{2}\,d{\bf v}_{2} \nonumber \\
&&- \sum_{\alpha'\neq \alpha}\int N_{\alpha'} m_{\alpha'}\frac{\partial u_{i2}}
{\partial {\bf r}_{i}}
P_{2}^{(\alpha\alpha')}({\bf r}_{i},{\bf v}_{i},{\bf r}_{2},{\bf v}_{2},t)\,
d{\bf r}_{2}\,d{\bf v}_{2}.
\end{eqnarray}
From the $N$-body Fokker-Planck equation (\ref{eq.fokker}), we can obtain an
equation for the time evolution of the two-body distribution function
$P_{2}^{(\alpha\alpha')}({\bf r}_{i},{\bf v}_{j},{\bf r}_{j},{\bf v}_{j},t)$ 
and again show that, in the proper thermodynamic limit, the mean-field 
approximation (\ref{eq.meanfield}) becomes exact. In that case, the 
expression of $I$ simplifies into
\begin{equation}
\label{eq.I}
I = - P_{1}^{(\alpha)}({\bf r}_{i},{\bf v}_{i},t)
\int \sum_{\alpha'}N_{\alpha'}m_{\alpha'}P_{1}^{(\alpha')}({\bf r}_{2},{\bf v}_{2},t)
\frac{\partial u_{i2}}{\partial {\bf r}_{i}}\,d{\bf r}_{2}\,d{\bf v}_{2}
= P_{1}^{(\alpha)}({\bf r}_{i},{\bf v}_{i},t)\langle {\bf F}\rangle_{i},
\end{equation}
where $\left<\textbf{F}\right>_i = - \nabla_{i} \Phi$ is the mean force (by 
unit of mass) acting on particle $i$. Introducing the distribution function 
\begin{equation}
\label{df}
f_{\alpha}({\bf r},{\bf v},t) = N_{\alpha}m_{\alpha}P_{1}^{(\alpha)}({\bf r},{\bf v},t),
\end{equation}
the mean-field Fokker-Planck equation takes the form
\begin{equation}
\frac{\partial f_{\alpha}}{\partial t}
+ \textbf{v}\cdot \frac{\partial f_{\alpha}}{\partial \textbf{r}}
+ \left<\textbf{F}\right> \cdot \frac{\partial f_{\alpha}}{\partial \textbf{v}}
= \frac{\partial}{\partial \textbf{v}}\cdot\left(D_{\alpha}\frac{\partial f_{\alpha}}
{\partial \textbf{v}} + \xi_\alpha f_{\alpha}\textbf{v}\right).
\end{equation}

In the strong friction limit, the stochastic equations of motion are given by
\eqn{(\ref{eq.stocha2})}. In that case, the $N$-body Fokker-Planck equation reads
\begin{equation}
\label{eq.fokker2}
\frac{\partial P_N}{\partial t}
= \sum_{i = 1}^N \frac{\partial}{\partial \textbf{r}_i} \cdot \left(D'_{i}
\frac{\partial P_N}{\partial \textbf{r}_i} + \mu_{i}{P_N} \nabla_{i}U\right).
\end{equation}
We note that the stationary solution of this equation is given by the configurational
part of the canonical distribution (\ref{eq.PNrv}) provided that the diffusion coefficient
and the mobility are related to each other by the Einstein relation
\begin{equation}
\label{eq.eni}
D'_{i}=\frac{\mu_{i}}{\beta}.
\end{equation}
Assuming that $i\in I_{\alpha}$ and integrating over the other variables, we get
\begin{equation}
\label{eq.1bfokker2}
\frac{\partial P_1^{(\alpha)}}{\partial t}
= \frac{\partial}{\partial \textbf{r}_i} \cdot \left(D'_{\alpha}
\frac{\partial P_1^{(\alpha)}}{\partial \textbf{r}_i} + \mu_{\alpha}
\int P_N \nabla_{i}U \prod_{j\neq i} d{\bf r}_k\right).
\end{equation}
Evaluating the last term in the mean-field approximation as done previously, we find
that
\begin{equation}
\label{eq.1bfokker3}
\frac{\partial P_1^{(\alpha)}}{\partial t}
= \frac{\partial}{\partial \textbf{r}_i} \cdot \left(D'_{\alpha}
\frac{\partial P_1^{(\alpha)}}{\partial \textbf{r}_i}
+ \mu_{\alpha} m_{\alpha} P_1^{(\alpha)} \nabla_{i}\Phi \right),
\end{equation}
which is clearly the same as
\begin{equation}
\frac{\partial \rho_{\alpha}}{\partial t} = \frac{1}{\xi_{\alpha}}
\nabla \cdot \left(\frac{k_B T}{m_{\alpha}} \nabla \rho_{\alpha} +
\rho_{\alpha} \nabla \Phi\right).
\end{equation}

\end{section}

\begin{section}{Estimate of the critical temperature using the Jeans swindle}
\label{app.jeans}

In this Appendix, we extend the original Jeans instability criterion
to the case of a multi-components system. We make the Jeans swindle,
assuming that the unperturbed state is infinite and homogeneous. Then,
we use this criterion to obtain an estimate of the critical temperature 
$T_{c}$ of an inhomogeneous isothermal multi-components
self-gravitating system confined within a box. 

Let us consider a small
perturbation around an equilibrium state of the two-components
Smoluchowski-Poisson system. The linearized equations for the
perturbation can be written
\begin{equation}
\label{eq.perturb1}
\begin{array}{rcl}
\displaystyle \frac{\partial \delta \rho_1}{\partial t} &=&
\displaystyle \frac{1}{\xi_1} \nabla \cdot \left(\frac{k_B T}
{m_1} \nabla \delta\rho_1 + \rho_{1} \nabla \delta \Phi
+ \delta \rho_1 \nabla \Phi\right),%
\vspace{6pt} \\
\displaystyle \frac{\partial \delta \rho_2}{\partial t} &=&
\displaystyle \frac{1}{\xi_2} \nabla \cdot \left(\frac{k_B T}
{m_2} \nabla \delta\rho_2 + \rho_{2} \nabla \delta \Phi
+ \delta \rho_2 \nabla \Phi\right),
\end{array}
\end{equation}
where $\rho$ and $\Phi$ refer to the equilibrium state. They have 
to be completed with the linearized Poisson equation
\begin{eqnarray}
\label{eq.poispert}
\Delta \delta\Phi =
S_d G \delta\rho.
\end{eqnarray}
These equations are exact but they remain complicated if the static
solution is inhomogeneous. They can be solved (semi-analytically) for
a one-component system \cite{c} but the generalization to a 
multi-components system is not straightforward. We shall invoke here the
Jeans swindle and consider that the equilibrium state is infinite and
homogeneous although this does not rigorously satisfy the equations at
zeroth order. With this simplifying assumption, using
\eqn{(\ref{eq.poispert})}, the linearized equations
(\ref{eq.perturb1}) take the form
\begin{equation}
\label{eq.perturb2}
\begin{array}{rcl}
\displaystyle \frac{\partial \delta \rho_1}{\partial t} &=&
\displaystyle \frac{1}{\xi_1} \left(\frac{k_B T}{m_1} \Delta
\delta\rho_1 + \rho_1 S_{d}G   \delta\rho \right),%
\vspace{6pt} \\
\displaystyle \frac{\partial \delta \rho_2}{\partial t} &=&
\displaystyle \frac{1}{\xi_2} \left(\frac{k_B T}{m_2} \Delta
\delta\rho_2 + \rho_{2} S_{d}G   \delta\rho \right).
\end{array}
\end{equation}
Writing the perturbation as $\delta \rho_{\alpha} \sim
e^{i(\textbf{k} \cdot \textbf{r} - \omega t)}$, we get
\begin{equation}
\label{eq.perturb3}
\begin{array}{rcccccc}
\displaystyle \left( -i \omega\xi_{1} 
+\frac{k_B T}{m_1} k^2 - S_d G \rho_{1}\right) 
\delta \rho_1 - {S_d G} \rho_{1} \delta \rho_2=0,
\vspace{6pt} \\
\displaystyle  - {S_d G} \rho_{2} \delta \rho_1+\left(-i\omega\xi_{2} 
+\frac{k_B T}{m_2} k^2 - S_d G \rho_{2}\right) 
\delta \rho_2=0.
\end{array}
\end{equation}
The cancellation of the determinant of the above system determines the
dispersion relation. One can show that $\gamma=-i\omega$ is real so
it represents either the growth rate of the perturbation ($\gamma>0$)
or its exponential damping ($\gamma<0$)
\footnote{Note that if we start from the two-components barotropic Euler 
equations (which are the usual equations used in Jeans analysis)
instead of the two-components Smoluchowski equations, we get the same
form of dispersion relation except that $i\omega\xi_{\alpha}$ is
replaced by $\omega^{2}$ and $k_{B}T/m_{\alpha}$ is replaced by the
velocity of sound $c_{\alpha}^{2}$. In that case $\omega^{2}$ is real
; for $\omega^{2}>0$, the system is stable and the perturbation
oscillates with pulsation $\omega$ and for $\omega^{2}<0$ the system
is unstable and the growth rate is $|i\omega|$.}. The point of
marginal stability $\omega=0$ is obtained for the Jeans wavevector
\begin{equation}
\label{eq.k2j}
k^2_J = \frac{S_d G  \left(m_1\rho_{1}+m_2\rho_{2}\right)}{k_B T}.
\end{equation}
More generally, for a multicomponent system, we have found that
\begin{equation}
\label{eq.k2jb}
k^2_J = \frac{S_d G}{ k_B T}  \sum_{\alpha=1}^{X} m_\alpha\rho_{\alpha}.
\end{equation}
The criterion of instability $-i \omega
\ge 0$ is equivalent to $k\le k_J$. In terms of the wavelength $\lambda=2\pi/k$, 
it can be written
\begin{equation}
\label{eq.lj}
\lambda^2 \ge \lambda^2_J
= \frac{4 \pi^2 k_{B}T}{S_d G \left(m_{1}\rho_{1}
+m_{2} \rho_{2}\right)}
\end{equation}
This criterion means that if the size of the perturbation $\lambda$ is
larger than the critical value $\lambda_J$, the gravitational
attraction will prevail over diffusion and the system will collapse. 
If we now return to our original problem, namely an isothermal system 
enclosed within a box of radius $R$, a na\"{\i}ve application of the 
above criterion indicates that the system is unstable if $R>\lambda_{J}$. 
Introducing the total mass of each species through the relation 
$\rho_{\alpha}\sim M_{\alpha}/R^{d}$, this criterion can be rewritten in 
terms of the temperature as
\begin{equation}
\label{eq.tcw}
T \le T_{c}\equiv K_{d}\frac{G(m_{1}M_{1}+m_{2}M_{2})R^{2-d}}{k_{B}}.
\end{equation}
As noticed in \cite{c}, the critical temperature $T_{c}$ marking the
gravitational instability of box-confined isothermal systems can be
related to the Jeans instability criterion by the above argument. Of
course, this na\"{\i}ve approach cannot catch the numerical constant
$K_{d}$ which appears in the expression of the critical
temperature. However, this constant can be obtained from the numerical
study of the single species case in $d=3$ where we have
$T_{c}=GMm/2.52 k_{B}R$
\cite{c}. Thus, we take $K_{3}=1/2.52$. Now, the interest of our
treatment for a multi-components system is that we can obtain the 
dependence of the critical temperature with $\mu$ and $\chi$. 
Returning to dimensionless variables, we get the instability criterion
\begin{equation}
\label{eq.jeans}
\eta \ge \eta_J = 2.52\frac{1 + \chi}{1 + \chi \mu},
\end{equation}
where $\eta_J$ is the critical inverse temperature obtained by using
the Jeans swindle. This expression returns the single-species result
for $\chi=0$ and for $\mu=1$. It is also consistent with the
single-species result for $\chi\rightarrow +\infty$ if we redefine
$\eta$ with $m_{1}$ instead of $m_{2}$. If we consider the limit
$\mu\rightarrow 0$ or $+\infty$ with fixed $N_{1}/N_{2}$, then
$\chi=(N_1/N_2)\mu\rightarrow 0$ or $+\infty$, and we again recover the
single species case. More generally, we see in \fig{\ref{fig.eclc3d}}
that this approximate expression gives a fair agreement with the exact
solution. This is quite satisfying in view of the approximations made
to arrive at \eqn{(\ref{eq.jeans})} (we have assumed that the system is
homogeneous). The relative success of this na\"{\i}ve approach is explained
by the fact that in $d=3$ the system is weakly inhomogeneous at
$T_{c}$. By contrast, the expression (\ref{eq.jeans}) does not work at
all in $d=2$ (compare with the exact values (\ref{eq.etac1}) and
(\ref{eq.etac2})) because the system tends to a Dirac peak at $T=T_c$.

\end{section}



\begin{thebibliography}{25}

\bibitem{prs} P.-H.~Chavanis, C.~Rosier, C.~Sire, Phys.~Rev.~E \textbf{66}, 036105 (2002).
\bibitem{sc} C.~Sire, P.-H.~Chavanis, Phys.~Rev.~E \textbf{66}, 046133 (2002).
\bibitem{lang} P.-H.~Chavanis, C.~Sire, Phys.~Rev.~E \textbf{69}, 016116 (2004).
\bibitem{sic} C.~Sire, P.-H.~Chavanis, Phys.~Rev.~E \textbf{69}, 066109 (2004).
\bibitem{chs} P.-H.~Chavanis, C.~Sire, Phys.~Rev.~E \textbf{70}, 026115 (2004).
\bibitem{crrs} P.-H.~Chavanis, M.~Ribot, C.~Rosier, C.~Sire, Banach Center Publ. {\bf 66}, 103 (2004).
\bibitem{sich} C.~Sire, P.-H.~Chavanis, Banach Center Publ. {\bf 66}, 287 (2004).
\bibitem{chav} P.-H.~Chavanis, (2004) [condmat/0409641].
\bibitem{planetes} P.-H.~Chavanis,  A\&A \textbf{356}, 1089 (2000).
\bibitem{csr} P.-H.~Chavanis, J. Sommeria, R. Robert, Astrophys. J. \textbf{471}, 385 (1996).
\bibitem{kurihouches} G. Kurizki, S. Giovanazzi, D. O'Dell, A.I. Artemiev, {\it New regimes 
in cold gases via laser-induced long-range interactions}, in Dynamics and Thermodynamics of 
Systems with Long Range Interactions, edited by T. Dauxois, S. Ruffo, E. Arimondo, M. Wilkens, 
Lecture Notes in Physics Vol. {\bf 602}, Springer (2002).
\bibitem{od} G.~Kurizki, Int. J. Mod. Phys. B \textbf{18}, 2027 (2004)
\bibitem{c} P.-H.~Chavanis, A~\&~A \textbf{381}, 340 (2002).
\bibitem{fermions} P.-H.~Chavanis, Phys. Rev. E  \textbf{65}, 056123 (2002).
\bibitem{ks} E.~Keller, L.A.~Segel, J.~theor.~Biol.~ \textbf{26}, 399 (1970).
\bibitem{jager}  W. J\"ager, S. Luckhaus, Trans. Am. Math. Soc. {\bf 329}, 819 (1992).
\bibitem{murray}  {\small J.D. Murray, {\it Mathematical Biology}
(Springer, 1991).}
\bibitem{katz}  J. Katz, Found. Phys. {\bf 33}, 223 (2003).
\bibitem{antonov} V.A. Antonov, Vest. Leningr. Gos. Univ. {\bf 7},
135 (1962).  
\bibitem{lbw} D. Lynden-Bell and R. Wood, Mon. Not. R. astr. Soc.
{\bf 138}, 495 (1968).
\bibitem{taf} L.~G.~Taff,~H.~M.~Van~Horn,~C.~J.~Hansen,~R.~R.~Ross,
Astrophys.~J. \textbf{197}, 651 (1975).
\bibitem{veg} H.~J.~De~Vega, J.~A.~Siebert, Phys.~Rev~E \textbf{66}, 016112 (2002).
\bibitem{mil} K.~R.~Yawn, B.~N.~Miller, Phys.~Rev.~E \textbf{68}, 056120 (2003).
\bibitem{kandrup} H. Kandrup, Astrophys. J. \textbf{244}, 316 (1981).
\bibitem{kinA} P.-H.~Chavanis, Physica A \textbf{332}, 89 (2004).
\bibitem{cll} P.-H.~Chavanis, P.~Lauren\c cot, M.~Lemou, Physica~A \textbf{341}, 145 (2004).
\bibitem{fp} P.-H.~Chavanis, Phys. Rev. E \textbf{68}, 036108 (2003).
\bibitem{w} G.~Wolansky, European~J.~Appl.~Math. \textbf{13}, 641 (2002).
\bibitem{metastable} P.-H.~Chavanis, A\&A, \textbf{432}, 117 (2005).
\bibitem{paddy}  {\small T. Padmanabhan, Phys. Rep.  {\bf 188}, 285 (1990).}
\bibitem{grand} P.-H.~Chavanis, A~\&~A \textbf{401}, 15 (2003).
\bibitem{bt}  {\small J. Binney and S. Tremaine,
{\it Galactic Dynamics} (Princeton Series in Astrophysics, 1987).}


\end{thebibliography}
\end{document}